\documentclass[12pt]{article}
\usepackage{amsmath,amssymb,amsthm}
\usepackage{graphicx}

\newcommand{\eqdef}{\stackrel{\text{def}}{=}}
\newcommand{\n}{\nonumber\\}
\newcommand{\bm}{\boldsymbol}
\numberwithin{equation}{section}
\newcommand{\ai}{\text{I}}
\newcommand{\ait}{\text{II}}
\newcommand{\Romannumeral}[1]{\uppercase\expandafter{\romannumeral#1}}
\newcommand{\cF}{c_{\text{\tiny$\mathcal{F}$}}}
\newcommand{\I}{\text{\Romannumeral{1}}}
\newcommand{\II}{\text{\Romannumeral{2}}}

\newtheorem*{thm}{\bf Theorem}

\newcommand{\ma}{\hspace{0pt}}

\allowdisplaybreaks[4]

\begin{document}

\baselineskip=20pt

\newfont{\elevenmib}{cmmib10 scaled\magstep1}
\newcommand{\Title}[1]{{\baselineskip=26pt
   \begin{center} \Large \bf #1 \\ \ \\ \end{center}}}
\newcommand{\Author}{\begin{center}
   \large \bf  Ryu Sasaki \end{center}}
\newcommand{\Address}{\begin{center}
      Department of Physics, Shinshu University,
     Matsumoto 390-8621, Japan\\
     Department of Physics, National Taiwan University, 
     Taipei 10617, Taiwan
   \end{center}}
\newcommand{\Accepted}[1]{\begin{center}
   {\large \sf #1}\\ \vspace{1mm}{\small \sf Accepted for Publication}
   \end{center}}
\newcommand{\Published}{\begin{center}
   {\small \sf Published in ``The Universe" {\bf 2} (2014) No.2 2-32.}
   \end{center}}

\thispagestyle{empty}

\Title{Exactly Solvable Quantum Mechanics}

\Author

\Address
\vspace{1cm}

\begin{abstract}
A comprehensive review of exactly solvable quantum mechanics  
is presented with the
emphasis of the recently discovered multi-indexed orthogonal polynomials. 
 The main subjects
to be discussed are the factorised Hamiltonians, the general structure of
the solution spaces of the Schr\"{o}dinger equation (Crum's theorem and its
modifications), the shape invariance, the exact solvability in the
Schr\"{o}dinger picture as well as in the Heisenberg picture, the
creation/annihilation operators and the dynamical symmetry algebras,
coherent states, various deformation schemes (multiple Darboux transformations) 
and the infinite families of multi-indexed orthogonal  polynomials, 
the exceptional orthogonal polynomials, and deformed 
exactly solvable scattering problems. 
\end{abstract}

\Published

{\bf PACS}: 03.65.-w, 03.65.Ca, 03.65.Fd, 03.65.Ge, 03.65.Nk, 02.30.Ik, 02.30.Gp

{\bf keywords}:
{exact solvability, 
factorised Hamiltonian, intertwining relations,
 shape invariance, Heisenberg operator solutions, closure relations,
 new orthogonal polynomials, reflectionless potentials}
 
\bigskip

\goodbreak

\tableofcontents
\section{Introduction}
\label{intro}

Assuming the rudimentary knowledge of quantum mechanics \cite{dirac,landau}, 
we start with the {\em factorised} Hamiltonians \eqref{factHam} 
and the Schr\"{o}dinger equations \eqref{Sch_eq}.
The general structure of the solution spaces is explored by the
intertwining relations and Crum's theorem \eqref{crums1}--\eqref{crumwronsfin} 
together with its modifications
\eqref{genmodhamdef}--\eqref{modnorm}. 
The multiple Darboux transformations are discussed 
generically \eqref{scheq}--\eqref{Mwron}.
Exact solvability in the Schr\"{o}dinger picture is explained by the
{\em shape invariance} \eqref{shapeinv}.
The {\em generic eigenvalue formula} \eqref{shapeenergy}, 
{\em unified Rodrigues formulas} \eqref{phin=A..Aphi0}
and the {\em forward/backward shift operators} 
\eqref{Fdef}--\eqref{Bdef} are deduced.
The solvability in the Heisenberg picture is derived based on the
{\em closure relation} \eqref{closurerel} between the 
{\em sinusoidal coordinate} \eqref{etaHei}
and the Hamiltonian.
The {\em creation/annihilation} operators \eqref{apm}--\eqref{apmphi}  
are introduced and their connection
with the {\em three term recurrence relations} \eqref{threeterm} 
of the orthogonal polynomials is emphasised.
The {\em dynamical symmetry algebras} \eqref{[H,apm]}--\eqref{apamg}
generated by the Hamiltonian and the
creation/annihilation operators are also established for all the solvable
systems in the Heisenberg picture. 
New orthogonal polynomials  are constructed in \S\ref{sec:Exce}.  
The radial oscillator \eqref{ex2} and P\"oschl-Teller 
 \eqref{ex2} systems are rationally extended in terms 
 of the {\em virtual state wave functions\/}
  \eqref{vsL1}--\eqref{vsJ2}. 
  The {\em multi-indexed orthogonal polynomials\/} 
 \eqref{multiP}--\eqref{multiortho} are obtained, 
 which are generalisation of the
 {\em exceptional orthogonal polynomials\/} \eqref{excepts}. 
 The duality \eqref{poteq}--\eqref{eigeq2} between the 
 pseudo virtual states \eqref{pvsH}--\eqref{pvsJ} and the eigenstates
is demonstrated.
The next topic  is  exactly solvable scattering problems 
and their extensions in \S\ref{sec:Scat}.
After the definition of the {\em scattering amplitudes} 
\eqref{psik}--\eqref{half}, {\em reflectionless
potentials\/} \eqref{cjsign}--\eqref{reflplane} are 
derived from the trivial potential $U\equiv0$ 
by multiple Darboux transformations. 
{\em Multi-indexed scattering amplitudes\/} 
\eqref{trDfull}--\eqref{rBC} are 
expressed in terms of the {\em asymptotic exponents\/} 
\eqref{asymfulls}--\eqref{asymcouls} of the polynomial type seed solutions. 
As a typical example, various data of the soliton potential
\eqref{solpot}--\eqref{defrt} are presented.
 The final section is for a summary and comments.
Basic symbols,  definitions and some formulas are listed in Appendix.
 
We usually discuss three elementary examples of exactly solvable potentials,
the harmonic oscillator, the radial oscillator 
and the P\"oschl-Teller potentials,
which are indicated by the initial of the polynomials constituting 
the corresponding eigenfunctions, that is,  
the Hermite (H), the Laguerre (L) and the Jacobi (J) polynomials.

Due to the length constraints, we have to concentrate on the systems of
single degree of freedom, which are the most basic and best established
part of the theory.
We apologise to all the authors whose good  works could not be referred to
in the review due to the lack of space.

\section{General Formulation}
\label{genform}

For the general settings of quantum mechanics, 
we refer to the standard textbooks \cite{dirac, landau}.
In this review we discuss exactly solvable Schr\"odinger equations.
In other words we present various methods of constructing 
such potentials $U(x)\in\mathbf{R}$ for which
the eigenvalue problem of the Hamiltonian $\mathcal{H}$
\begin{equation*}
\mathcal{H}\psi(x)=\mathcal{E}\psi(x),\qquad
\mathcal{H}\eqdef \sum_j\frac{p_j^2}{2m_j}+U(x),
\end{equation*}
is exactly solvable.
We will concentrate on the most fundamental case, that is, 
the one dimensional quantum mechanics (1-d QM).
Generalisation to multi-degrees of freedom cases 
will be mentioned in appropriate places.

\subsection{Problem Setting:1-d QM}
\label{probset}

Let us consider one-dimensional QM defined in  an interval $(x_1,x_2)$, 
in which $x_1$ and/or $x_2$ can be infinite.
For finite $x_j$, $j=1,2$, 
the potential must provide an infinite barrier $\lim_{x\to x_j}U(x)=+\infty$
at that boundary
lest the particle {\em tunnel out} from $(x_1,x_2)$. 
This fact provides proper boundary conditions
of the wavefunctions.
The dynamical variables are the coordinate $x$ and
its conjugate momentum $p$, which is realised as a differential operator
$p=-i\hbar\frac{d}{dx}\equiv -i\hbar\partial_x$.
Hereafter we adopt the convention $\hbar=1$ and $2m=1$ 
and consider  the following Hamiltonian
\begin{equation}
\mathcal{H}\eqdef -\frac{d^2}{dx^2}+U(x),\qquad x_1<x<x_2,
\label{1dham}
\end{equation}
with a {\em smooth} potential $U(x)\in \mathbf{C}^\infty$. 
We also require that the Hamiltonian is {\em bounded from below}.
The {\em eigenvalue problem} is to find all the discrete 
eigenvalues $\{\mathcal{E}(n)\}$ and the corresponding 
eigenfunctions $\{\phi_n(x)\}$
\begin{equation}
\mathcal{H}\phi_n(x)=\mathcal{E}(n)\phi_n(x),\quad n=0,1,\ldots,
\label{Sch_eq}
\end{equation}
of the given Hamiltonian $\mathcal{H}$ \eqref{1dham}.
The numbering of the eigenvalues is monotonously increasing:
\begin{equation}
  \mathcal{E}(0)<\mathcal{E}(1)<\mathcal{E}(2)<\cdots.
  \label{increase}
\end{equation}
The eigenfunctions are mutually orthogonal
\begin{equation}
(\phi_n,\phi_m)\eqdef 
\int_{x_1}^{x_2}\!\phi_n(x)^*\phi_m(x)dx=h_n\delta_{n\,m},\quad 
0<h_n<\infty,\quad n,m=0,1,\ldots,
\end{equation}
which is a consequence of the {\em hermiticity} of the Hamiltonian 
$\mathcal{H}$, \cite{dirac,landau}.
Then the  {\em oscillation theorem} asserts that the 
{\em $n$-th eigenfunction $\phi_n(x)$ has $n$ 
simple zeros in $(x_1,x_2)$}. In particular the {\em ground state eigenfunction} 
$\phi_0(x)$ has no zero in $(x_1,x_2)$, 
and we will choose the convention that it is {\em positive} $\phi_0(x)>0$.
We also choose all the eigenfunctions to be {\em real}, 
$\phi_n(x)\in\mathbf{R}$.

Roughly speaking we encounter two types of problems. 
The first is {\em confining potentials} 
$\lim_{x\to x_1+0}U(x)=+\infty=\lim_{x\to x_2-0}U(x)$,
which has {\em infinitely many discrete eigenvalues}.
The rest is non-confining and it  has 
 {\em  finitely many or infinite%
 \footnote{The Coulomb potential is 
 a well-known example of non-confining potential 
 having infinitely many discrete eigenstates.} 
 discrete eigenvalues} and if $x_1=-\infty$ and/or $x_2=+\infty$,
{\em scattering problems} can be considered. 
The setting of scattering problems will be 
introduced at the beginning of section \ref{sec:Scat}.

When all the eigenvalues, finite or infinite in number, 
and the corresponding 
eigenfunctions can be obtained explicitly, 
such a potential is called  {\em exactly solvable} \cite{infhull,susyqm}.
There are some potentials for which  only  finitely many eigenvalues 
and eigenfunctions can be 
obtained explicitly. 
Such potentials are called 
{\em quasi-exactly solvable} \cite{morozov, turb, ush, st1}.

\subsection{Factorised Hamiltonian}
\label{sec:H}

Let us consider the eigenvalue problem of 
a given Hamiltonian \eqref{1dham}  having a finite 
or semi-infinite number of {\em discrete energy levels}.
The additive constant of the Hamiltonian is so chosen that the ground state
energy vanishes, $\mathcal{E}(0)=0$. 
That is, the Hamiltonian is {\em positive semi-definite}.
It is a well known theorem in linear algebra that any positive
semi-definite hermitian matrix can be factorised as a product of a certain
matrix, say $\mathcal{A}$, and its hermitian conjugate $\mathcal{A^\dagger}$.
As we will see shortly, the Hamiltonians we consider  always have
factorised forms in one-dimension as well as in higher dimensions.

The Hamiltonian we consider has a simple factorised form
\cite{infhull}
\begin{equation}
  \mathcal{H}\eqdef\mathcal{A}^\dagger \mathcal{A}\quad \text{or}\quad
  \mathcal{H}\eqdef\sum_{j=1}^D\mathcal{A}_j^\dagger\mathcal{A}_j\quad
  \text{in }D\ \text{dimensions}.
  \label{factHam}
\end{equation}
The operators $\mathcal{A}$ and $\mathcal{A}^\dagger$ in 1-d
QM are:
\begin{align}
&  \mathcal{A}\eqdef\frac{d}{dx}-\frac{dw(x)}{dx},\quad 
  \mathcal{A}^\dagger=-\frac{d}{dx}-\frac{dw(x)}{dx},\quad
  w(x)\in\mathbf{R},\quad\phi_0(x)=e^{w(x)},
  \label{oAdef}\\
  &\mathcal{H}=p^2+U(x),\quad
  U(x)\eqdef\bigl(\partial_x w(x)\bigr)^2+\partial_x^2w(x),
  \label{oUdef}
\end{align}
in which a real function $w(x)$ is called a {\em prepotential}.
The Hamiltonian of a multi-degrees of freedom   
system can be constructed in a similar way:
\begin{equation}
  \mathcal{A}_j\eqdef\frac{\partial}{\partial x_j}
  -\frac{\partial w(x)}{\partial x_j},
  \ \ \mathcal{A}_j^\dagger=-\frac{\partial}{\partial x_j}
  -\frac{\partial w(x)}{\partial x_j}\ \ (j=1,\ldots,D),\quad
  \phi_0(x)=e^{w(x)}.
  \label{oAdefj}
\end{equation}
The prepotential approach is also useful in Calogero-Moser systems
\cite{Cal-Sut} in QM \cite{bms,kps}.

\bigskip
The Schr\"{o}dinger equation \eqref{Sch_eq}
is a second order differential equation  and
the {\em ground state wavefunction} $\phi_0(x)$ is determined as a zero
mode of the operator $\mathcal{A}$ ($\mathcal{A}_j$) which is a first
order equation:
\begin{equation}
  \mathcal{A}\phi_0(x)=0\quad(\mathcal{A}_j\phi_0(x)=0,\ j=1,\ldots,D)\quad
  \Rightarrow\ \mathcal{H}\phi_0(x)=0.
  \label{Aphi0=0}
\end{equation}
It should be stressed that the inverse of the zero mode of $\mathcal{A}$ is the
zero mode of $\mathcal{A}^\dagger$:
\begin{equation}
  \mathcal{A}^\dagger\phi_0^{-1}(x)=0\quad
  (\mathcal{A}_j^\dagger\phi_0^{-1}(x)=0,\ j=1,\ldots,D).
  \label{Adagphi0=0}
\end{equation}
This fact simply means that a quasi-exactly solvable system 
with only one known eigenvalue $\mathcal{E}(0)=0$ 
and the corresponding eigenfunction $\phi_0(x)$ can always be constructed
by an arbitrary  positive and smooth square integrable function $\phi_0(x)$.

At the end of this subsection, let us emphasize 
that any non-vanishing (for example, positive)
solution of the original Hamiltonian \eqref{1dham}
\begin{equation}
\mathcal{H}\tilde{\phi}(x)=\tilde{\mathcal E}\tilde{\phi}(x),\quad 
\tilde{\mathcal E}\in\mathbf{R},
\quad \tilde{\phi}(x)>0,\quad x\in(x_1,x_2),
\label{nonzero}
\end{equation}
provides a non-singular factorisation of the original Hamiltonian
\begin{align} 
 \mathcal{H}=\tilde{\mathcal A}^\dagger 
 \tilde{\mathcal A}+\tilde{\mathcal E},\quad 
  \tilde{\mathcal A}\eqdef\frac{d}{dx}-
  \frac{\partial_x \tilde{\phi}(x)}{\tilde{\phi}(x)},\quad 
 \tilde{\mathcal A}^\dagger=
 -\frac{d}{dx}-\frac{\partial_x \tilde{\phi}(x)}{\tilde{\phi}(x)}.
\end{align}

\subsection{Intertwining Relations: Crum's Theorem}
\label{sec:Inter}

In this subsection we show the general structure of the solution space of 1-d QM.
Let us denote by $\mathcal{H}^{[0]}$ the
original factorised Hamiltonian $\mathcal{H}$ \eqref{1dham}, \eqref{factHam} 
and by $\mathcal{H}^{[1]}$ its {\em partner\/}
({\em associated\/}) Hamiltonian obtained by changing the order of
$\mathcal{A}^\dagger$ and $\mathcal{A}$:
\begin{equation}
 \mathcal{H}\equiv  \mathcal{H}^{[0]}\eqdef\mathcal{A}^\dagger\mathcal{A},\qquad
  \mathcal{H}^{[1]}\eqdef\mathcal{A}\mathcal{A}^\dagger.
  \label{factHam2}
\end{equation}
One simple and most important consequence of the factorised Hamiltonians
\eqref{factHam2} is the {\em intertwining relations\/}:
\begin{align}
  \mathcal{A}\mathcal{H}^{[0]}=\mathcal{A}\mathcal{A}^\dagger\mathcal{A}
  =\mathcal{H}^{[1]}\mathcal{A},\qquad
  \mathcal{A}^\dagger\mathcal{H}^{[1]}
  =\mathcal{A}^\dagger\mathcal{A}\mathcal{A}^\dagger
  =\mathcal{H}^{[0]}\mathcal{A}^\dagger.
  \label{intw1}
\end{align}
The pair of Hamiltonians $\mathcal{H}^{[0]}$ and $\mathcal{H}^{[1]}$ are
essentially {\em iso-spectral\/} and their eigenfunctions
$\{\phi_n^{[0]}(x)\}$ and $\{\phi_n^{[1]}(x)\}$ are related by the
Darboux-Crum transformations \cite{darboux,crum}:
\begin{align}
  &\mathcal{H}^{[0]}\phi_n^{[0]}(x)=\mathcal{E}(n)\phi_n^{[0]}(x)
  \quad(n=0,1,\ldots),\quad\mathcal{A}\phi_0^{[0]}(x)=0,\\
  &\mathcal{H}^{[1]}\phi_n^{[1]}(x)=\mathcal{E}(n)\phi_n^{[1]}(x)
  \quad(n=1,2,\ldots),\\
  &\phi_n^{[1]}(x)\eqdef\mathcal{A}\phi_n^{[0]}(x),\quad
  \phi_n^{[0]}(x)=\frac{\mathcal{A}^\dagger}{\mathcal{E}(n)}\phi_n^{[1]}(x)
  \quad(n=1,2,\ldots),
  \label{phi01}\\
  &(\phi_n^{[1]},\phi_m^{[1]})=
  \mathcal{E}(n)(\phi_n^{[0]},\phi_m^{[0]})
  \quad(n,m=1,2,\ldots).
\end{align}
The associated Hamiltonian $\mathcal{H}^{[1]}$, 
sometimes called the partner  Hamiltonian,
has the lowest eigenvalue
$\mathcal{E}(1)$ and the state corresponding to $\mathcal{E}(0)$ is missing.
In other words, the ground state corresponding to $\mathcal{E}(0)$ is {\em deleted}
by the transformation $\mathcal{H}^{[0]}\to \mathcal{H}^{[1]}$.

 If the ground state energy $\mathcal{E}(1)$ is subtracted
from the partner Hamiltonian $\mathcal{H}^{[1]}$, it is again positive
semi-definite and can be factorised in terms of new operators
$\mathcal{A}^{[1]}$ and $\mathcal{A}^{[1]\dagger}$:
\begin{equation}
  \mathcal{H}^{[1]}=\mathcal{A}^{[1]\dagger}\mathcal{A}^{[1]}
  +\mathcal{E}(1),\quad\mathcal{A}^{[1]}\phi_1^{[1]}(x)=0.
\end{equation}
By changing the orders of $\mathcal{A}^{[1]\dagger}$ and
$\mathcal{A}^{[1]}$, a new Hamiltonian $\mathcal{H}^{[2]}$ is defined:
\begin{equation}
  \mathcal{H}^{[2]}\eqdef\mathcal{A}^{[1]}\mathcal{A}^{[1]\dagger}
  +\mathcal{E}(1).
\end{equation}
These two Hamiltonians, $\mathcal{H}^{[1]}-\mathcal{E}(1)$ 
and $\mathcal{H}^{[2]}-\mathcal{E}(1)$,
 are intertwined by $\mathcal{A}^{[1]}$ and
$\mathcal{A}^{[1]\dagger}$:
\begin{align}
  &\mathcal{A}^{[1]}\bigl(\mathcal{H}^{[1]}-\mathcal{E}(1)\bigr)
  =\mathcal{A}^{[1]}\mathcal{A}^{[1]\dagger}\mathcal{A}^{[1]}
  =\bigl(\mathcal{H}^{[2]}-\mathcal{E}(1)\bigr)\mathcal{A}^{[1]},
  \label{intw12}\\
  &\mathcal{A}^{[1]\dagger}\bigl(\mathcal{H}^{[2]}-\mathcal{E}(1)\bigr)
  =\mathcal{A}^{[1]\dagger}\mathcal{A}^{[1]}\mathcal{A}^{[1]\dagger}
  =\bigl(\mathcal{H}^{[1]}-\mathcal{E}(1)\bigr)\mathcal{A}^{[1]\dagger}.
  \label{intw21}
\end{align}
The iso-spectrality of the two Hamiltonians $\mathcal{H}^{[1]}$ and
$\mathcal{H}^{[2]}$ and the relationship among their eigenfunctions
follow as before:
\begin{align}
  &\mathcal{H}^{[2]}\phi_n^{[2]}(x)=\mathcal{E}(n)\phi_n^{[2]}(x)
  \quad(n=2,3,\ldots),\\
  &\phi_n^{[2]}(x)\eqdef\mathcal{A}^{[1]}\phi_n^{[1]}(x),\quad
  \phi_n^{[1]}(x)=\frac{\mathcal{A}^{[1]\dagger}}
  {\mathcal{E}(n)-\mathcal{E}(1)}\phi_n^{[2]}(x)
  \quad(n=2,3,\ldots),
  \label{phi12}\\
  &(\phi_n^{[2]},\phi_m^{[2]})=
  \bigl(\mathcal{E}(n)-\mathcal{E}(1)\bigr)(\phi_n^{[1]},\phi_m^{[1]})
  \quad(n,m=2,3,\ldots),\\
  &\mathcal{H}^{[2]}=\mathcal{A}^{[2]\dagger}\mathcal{A}^{[2]}
  +\mathcal{E}(2),\quad\mathcal{A}^{[2]}\phi_2^{[2]}(x)=0.
\end{align}
By the transformation $\mathcal{H}^{[1]}\to \mathcal{H}^{[2]}$ 
the state corresponding to 
$\mathcal{E}(1)$ is deleted.

This process can go on indefinitely by successively deleting the lowest
lying energy level:
\begin{align}
  &\mathcal{H}^{[s]}
  \eqdef\mathcal{A}^{[s-1]}\mathcal{A}^{[s-1]\dagger}+\mathcal{E}(s-1)
  =\mathcal{A}^{[s]\dagger}\mathcal{A}^{[s]}+\mathcal{E}(s),
  \label{crums1}\\
  &\mathcal{H}^{[s]}\phi_n^{[s]}(x)=\mathcal{E}(n)\phi_n^{[s]}(x)
  \quad(n=s,s+1,\ldots),\quad\mathcal{A}^{[s]}\phi_s^{[s]}(x)=0,\\
  &\phi_n^{[s]}(x)\eqdef\mathcal{A}^{[s-1]}\phi_n^{[s-1]}(x),\quad
  \phi_n^{[s-1]}(x)=\frac{\mathcal{A}^{[s-1]\dagger}}
  {\mathcal{E}(n)-\mathcal{E}(s-1)}\phi_n^{[s]}(x)
  \quad(n=s,s+1,\ldots),
  \label{phiss1}\\
  &(\phi_n^{[s]},\phi_m^{[s]})=
  \bigl(\mathcal{E}(n)-\mathcal{E}(s-1)\bigr)(\phi_n^{[s-1]},\phi_m^{[s-1]})
  \quad(n,m=s,s+1,\ldots).
\end{align}
The quantities in the $s$-th step are defined by those in the ($s-1$)-st
step: ($s\geq 1$)
\begin{align}
&w^{[s]}(x)\eqdef\log|\phi^{[s]}_s(x)|,\quad
\phi_n^{[s]}(x)\eqdef\mathcal{A}^{[s-1]}\phi_n^{[s-1]}(x),
  \label{w[s]}\\
  &\mathcal{A}^{[s]}\eqdef\partial_x-\partial_xw^{[s]}(x),\quad
  \mathcal{A}^{[s]\,\dagger}=-\partial_x-\partial_xw^{[s]}(x),
  \end{align}
The eigenfunctions at the $s$-th step have succinct {\em determinant forms}
in terms of the Wronskian: ($n\geq s\geq 0$)
\begin{align}
&\text{W}\,[f_1,\ldots,f_m](x)
  \eqdef\det\Bigl(\frac{d^{j-1}f_k(x)}{dx^{j-1}}\Bigr)_{1\leq j,k\leq m}
  \quad(\text{Wronskian}),
  \label{wron}\\
& \mathcal{H}^{[s]}=\mathcal{H}^{[0]}-2\partial_x^2
\log| {\text{W}\,[\phi_0,\phi_1,\ldots,\phi_{s-1}](x)}|,
\label{wronham}\\
  &\phi^{[s]}_n(x)=
  \frac{\text{W}\,[\phi_0,\phi_1,\ldots,\phi_{s-1},\phi_n](x)}
  {\text{W}\,[\phi_0,\phi_1,\ldots,\phi_{s-1}](x)},
  \label{crumwronsfin}
  \end{align}
In deriving the determinant formulas \eqref{wronham} and \eqref{crumwronsfin}
use is made of the following properties of the Wronskian
\begin{align}
  &\text{W}[gf_1,gf_2,\ldots,gf_n](x)
  =g(x)^n\text{W}[f_1,f_2,\ldots,f_n](x),
  \label{Wformula1}\\
  &\text{W}\bigl[\text{W}[f_1,f_2,\ldots,f_n,g],
  \text{W}[f_1,f_2,\ldots,f_n,h]\,\bigr](x)\n
  &=\text{W}[f_1,f_2,\ldots,f_n](x)\,
  \text{W}[f_1,f_2,\ldots,f_n,g,h](x)
  \qquad(n\geq 0).
  \label{Wformula2}
\end{align}
Another useful property of the Wronskian is that it is invariant when
the derivative $\frac{d}{dx}$ is replaced by an arbitrary `covariant
derivative' $D_i$ with an arbitrary smooth function $q_i(x)$:
\begin{equation}
  D_i\eqdef\frac{d}{dx}-q_i(x),\quad
  \text{W}[f_1,f_2,\ldots,f_n](x)=
  \det\bigl(D_{j-1}\cdots D_2D_1f_k(x)\bigr)_{1\le j,k\le n},
  \label{covW}
\end{equation}
with $D_{j-1}\cdots D_2D_1\bigm|_{j=1}=1$.
The norm of the $s$-th step eigenfunctions have a simple uniform expression:
\begin{equation}
  (\phi^{[s]}_n,\phi^{[s]}_m)
  =\prod_{j=0}^{s-1}\bigl(\mathcal{E}(n)-\mathcal{E}(j)\bigr)
  \cdot(\phi_n,\phi_m).
\end{equation}
To sum up, we have the following
\begin{thm}{\rm (Crum \cite{crum})}
For a given Hamiltonian system $\mathcal{H}\equiv\mathcal{H}^{[0]}$, there are
associated Hamiltonian systems $\mathcal{H}^{[1]}$, $\mathcal{H}^{[2]}$, \ldots,
as many as the total number of discrete eigenvalues 
of the original system $\mathcal{H}^{[0]}$.
They share the same eigenvalues $\{\mathcal{E}(n)\}$ 
of the original system and the eigenfunctions of  
 $\mathcal{H}^{[j]}$ and  $\mathcal{H}^{[j+1]}$ 
 are related linearly by $\mathcal{A}^{[j]}$ and 
 $\mathcal{A}^{[j]\dagger}$.
\end{thm}
This situation of the Crum's theorem is illustrated in Fig.\,1.
If the original system $\mathcal{H}^{[0]}$ is {\em exactly solvable}, 
then all the associated 
systems $\{\mathcal{H}^{[j]}\}$ are also  {\em exactly solvable}.

\bigskip

A quantum mechanical system with a factorised Hamiltonian
$\mathcal{H}=\mathcal{A}^\dagger\mathcal{A}$ together with the associated
one $\mathcal{H}^{[1]}=\mathcal{A}\mathcal{A}^\dagger$ is sometimes called
a `supersymmetric' QM \cite{witten,susyqm}. This seems to be  rather a misnomer,
since as we have shown the factorised form is generic and it implies no
extra symmetry. The iso-spectrality is shared by all the associated
Hamiltonians, not merely by the first two. In this connection, the
transformation of mapping the $s$-th to the ($s+1$)-st associated Hamiltonian
is sometimes called susy transformation. It is also known as the 
 Darboux-Crum transformation. Those covering multi-steps are sometimes
called `higher derivative' or `nonlinear' or `$\mathcal{N}$-fold' susy
transformations \cite{ais,plyu,aoy}.
We will discuss generic Darboux transformations in \S\ref{sec:darb}.

\begin{center}
  \includegraphics{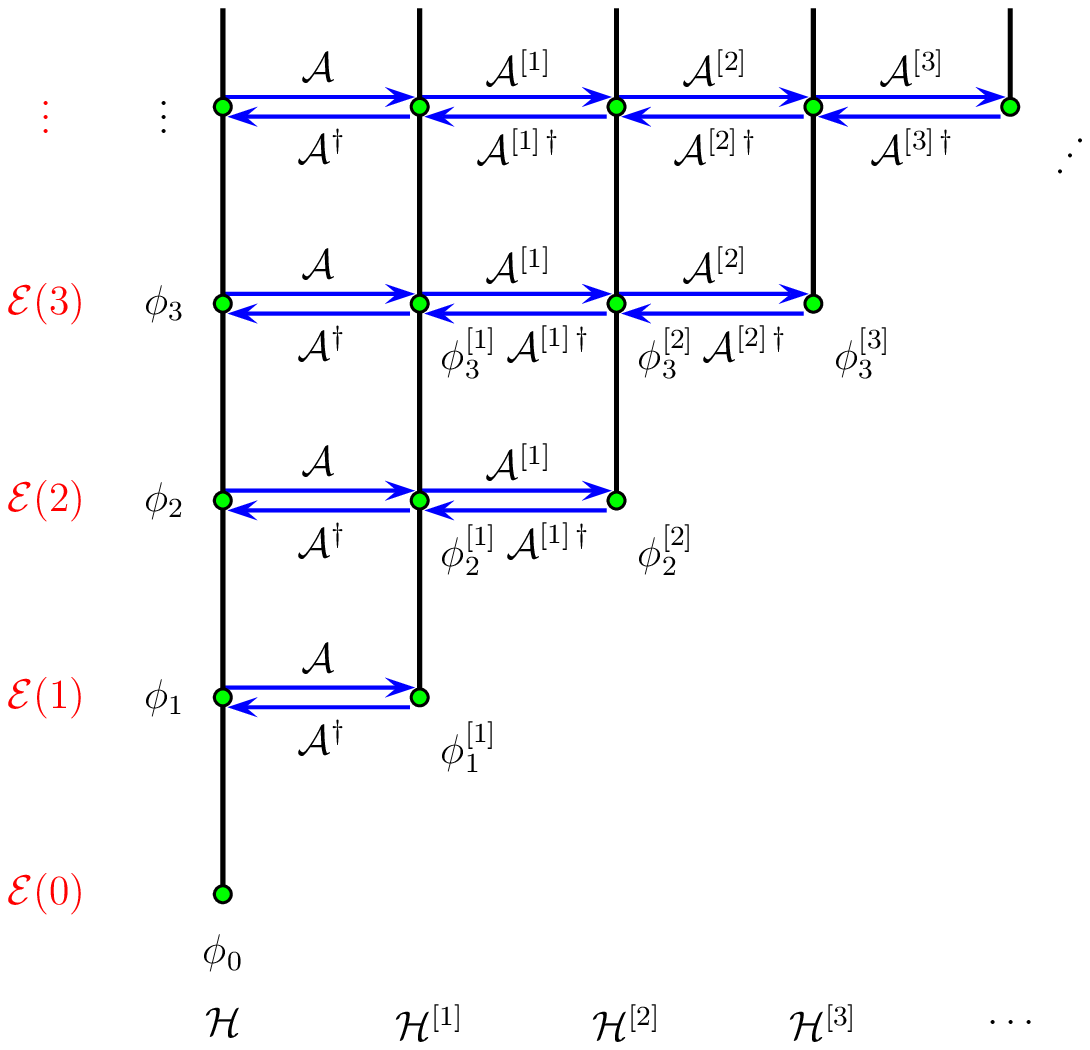}
\end{center}
\begin{center}
  Figure\,1: General structure of the solution space of 1-d QM.
\end{center}

\subsection{Modified Crum's Theorem}
\label{sec:ModCrum}

Crum's theorem provides a series of  essentially  
iso-spectral Hamiltonian systems by deleting
successively the  lowest lying eigenlevels from the original Hamiltonian system
$\mathcal{H}$ and $\{\phi_n(x)\}$.
The modification of Crum's theorem by Krein-Adler \cite{krein,adler} is
achieved by deleting a finite number of eigenstates indexed by a set of
distinct non-negative  integers
$\mathcal{D}\eqdef\{d_1,d_2,\ldots,d_{M}\}\subset\mathbb{Z}_{\ge0}^{M}$,
satisfying certain conditions to be specified later \eqref{adlercond}.
After the deletion, the new ground state has the label $\mu$ given by
\begin{equation}
  \mu\eqdef\min\{n\,|\,n\in\mathbb{Z}_{\geq 0}\backslash\mathcal{D}\}.
  \label{defmu}
\end{equation}
Corresponding to \eqref{w[s]}, the new essentially iso-spectral
Hamiltonian system is: %
\begin{align} 
  &\bar{\mathcal{H}}_{\mathcal D}\eqdef
  \bar{\mathcal{A}}_{\mathcal D}^\dagger\bar{\mathcal{A}}_{\mathcal D}+\mathcal{E}(\mu),
  \quad\bar{\mathcal{A}}_{\mathcal D}\,\bar{\phi}_{{\mathcal D};\mu}(x)=0,
  \label{genmodhamdef}\\
  &\bar{\mathcal{H}}_{\mathcal D}\bar{\phi}_{{\mathcal D};n}(x)=
  \mathcal{E}(n)\bar{\phi}_{{\mathcal D};n}(x)
   \quad(n\in\mathbb{Z}_{\geq 0}\backslash\mathcal{D}),\\
&\bar{\mathcal{H}}_{\mathcal D}=p^2+\bar{U}_{\mathcal D}(x),\quad
  \bar{U}_{\mathcal D}(x)\eqdef U(x)-2\partial_x^2\Bigl(
  \log|\text{W}\,[\phi_{d_1},\phi_{d_2},\ldots,\phi_{d_{M}}](x)|\Bigr),
  \label{Ub1bs}\\
  &\bar{\mathcal{A}}_{\mathcal D}\eqdef
  \partial_x-\partial_x\bar{w}_{\mathcal D}(x),\quad
  \bar{\mathcal{A}}_{\mathcal D}^{\dagger}=
  -\partial_x-\partial_x\bar{w}_{\mathcal D}(x),\quad
  \bar{w}_{\mathcal D}(x)\eqdef\log|\bar{\phi}_{\mathcal D;\mu}(x)|,\\
  &\bar{\phi}_{{\mathcal D};n}(x)\eqdef
  \frac{\text{W}\,[\phi_{d_1},\phi_{d_2},\ldots,\phi_{d_{M}},\phi_n](x)}
  {\text{W}\,[\phi_{d_1},\phi_{d_2},\ldots,\phi_{d_{M}}](x)},
  \label{adlphin}
  \end{align}
and 
\begin{equation}
(\bar{\phi}_n,\bar{\phi}_m)
  =\prod_{j=1}^{\ell}\bigl(\mathcal{E}(n)-\mathcal{E}(d_j)\bigr)
  \cdot(\phi_n,\phi_m)\quad(n,m\in\mathbb{Z}_{\geq 0}\backslash\mathcal{D}).
  \label{modnorm}
\end{equation}
It should be emphasised that the Hamiltonian $\bar{\mathcal{H}}$ as well as
the eigenfunctions $\{\bar{\phi}_n(x)\}$ are symmetric with respect to
$d_1,\ldots,d_{M}$, and thus they are independent of the order of $\{d_j\}$.
In order to guarantee  the positivity of the norm \eqref{modnorm} of all
the eigenfunctions $\{\bar{\phi}_n(x)\}$ of the  modified Hamiltonian,
the set of deleted energy levels $\mathcal{D}=\{d_1,\ldots,d_{M}\}$ must
satisfy the necessary and sufficient conditions \cite{krein,adler}
\begin{equation}
  \prod_{j=1}^{M}(m-d_j)\ge0,\quad \forall m\in\mathbb{Z}_{\geq 0}.
  \label{adlercond}
\end{equation}
And the Hamiltonian $\bar{\mathcal{H}}$ 
and the potential $\bar{U}(x)$ \eqref{Ub1bs}  
is non-singular under this condition (see \cite{adler} for details).
 Crum's theorem in \S\,\ref{sec:Inter} corresponds to the choice
$\{d_1,d_2,\ldots,d_{M}\}=\{0,1,\ldots,M-1\}$.
The algebraic derivation of the formulas \eqref{genmodhamdef}--\eqref{modnorm}
is essentially the same as that of Crum's theorem.
See Fig.2 for the illustration of the Krein-Adler
transformation.

\begin{center}
  \includegraphics{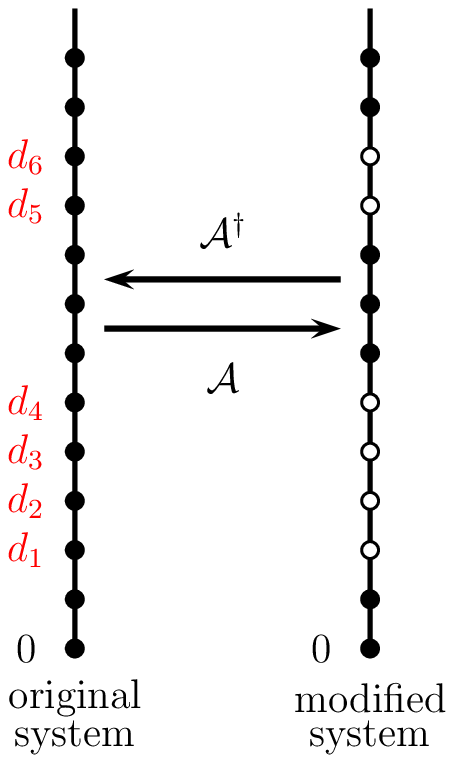}
\end{center}
{\baselineskip=14pt
\begin{quote}
  Figure\,2: General image of the Krein-Adler transformation. 
  The black circles denote the energy levels, whereas the white circles
denote {\em deleted\/} energy levels. 
\end{quote}
}

Starting from an {\em exactly solvable} Hamiltonian, 
one can construct infinitely
many variants of {\em exactly solvable} Hamiltonians 
and their eigenfunctions by
Krein-Adler transformations. The resulting systems are,
however, not shape invariant, even if the starting system is. 

\subsection{Darboux Transformations}
\label{sec:darb}

Darboux transformations \cite{darboux} in general apply 
in much wider contexts than 1-d QM,
{\em i.e,\/}  without the constraints on the boundary
conditions, self-adjointness or factorisation of $\mathcal{H}$, 
Hilbert space, etc.
In terms of  {\em seed solutions} of a Schr\"odinger type equation, 
they map solutions
of the same equation to those belonging to a different potential {\em iso-spectrally}. 
The potential $U(x)$ in \eqref{1dham} can be complex valued 
and the {\em seed solutions} $\{\varphi_j(x)\}$ 
need not be eigenfunctions. 
Let $\psi(x)$ and  $\{\varphi_j(x), \tilde{\mathcal{E}}_j\}$ 
$(j=1,\ldots,M)$ be  distinct solutions,
not necessarily square-integrable eigenfunctions,  of 
the Schr\"odinger type equation:
\begin{equation}
\mathcal{H}\psi(x)=\mathcal{E}\psi(x),
\quad  \mathcal{H}\varphi_j(x)
= \tilde{\mathcal{E}}_j\varphi_j(x),
\qquad  \mathcal{E}, \tilde{\mathcal{E}}_j\in\mathbb{C},
\quad j=1,\ldots,M.
\label{scheq}
\end{equation}
By picking up one seed solution, say, $\varphi_1$,  
we define new functions
\begin{align}
&{\psi}^{(1)}(x)\eqdef \frac{\text{W}[\varphi_1,\psi](x)}{\varphi_1(x)} 
\equiv \frac{\varphi_1(x) \psi '(x) -\varphi '_1(x) \psi (x)}{\varphi_1(x)} ,\\
 &  \varphi^{(1)}_{1}(x)\eqdef\frac1{\varphi_1(x)},\quad 
 \varphi^{(1)}_{k}(x)\eqdef  
 \frac{\text{W}[\varphi_1,\varphi_k](x)}{\varphi_1(x)},
\quad k=2,\ldots,M.
\end{align}
It is elementary to show that they are solutions of 
a new Schr\"odinger type equation 
with a deformed Hamiltonian ${\mathcal H}^{(1)}$
\begin{equation}
{\mathcal H}^{(1)}=-\frac{d^2}{dx^2}+{U}^{(1)}(x),\quad
{U}^{(1)}(x)\eqdef U(x)-2\frac{d^2\log|\varphi_1(x)|}{dx^2},
\end{equation}
with the same energies $\mathcal{E}$ and $\tilde{\mathcal{E}}_k$:
\begin{equation}
{\mathcal H}^{(1)}{\psi}^{(1)}(x)=\mathcal{E}{\psi}^{(1)}(x),\quad
{\mathcal H}^{(1)}{\varphi}^{(1)}_j(x)=
\tilde{\mathcal{E}}_j{\varphi}^{(1)}_j(x),\quad j=1,\ldots,M.
\label{newschr}
\end{equation}
By repeating these transformations $M$-times, we arrive at
\begin{thm} {\rm (Darboux \cite{darboux})}
Let $\psi(x) $ be a solution of the original Schr\"odinger type equation 
\begin{equation}
\mathcal{H}\psi(x)=\mathcal{E}\psi(x) .
\label{scheq3}
\end{equation}
Then the functions 
\begin{align}
&\psi^{(M)}(x)\eqdef\frac{\text{W}[\varphi_1,\ldots,\varphi_M,\psi](x)}
{\text{W}[\varphi_1,\ldots,\varphi_M](x)},
\label{psiM}\\
&\varphi^{(M)}_j(x)\eqdef
\frac{\text{W}[\varphi_1,\ldots,\breve{\varphi}_j,\ldots,\varphi_M](x)}
{\text{W}[\varphi_1,\ldots,\varphi_M](x)},\quad j=1,\ldots,M,
\label{varphiM}
\end{align}
satisfy the $M$-th deformed Schr\"odinger type equation 
with the same energy:
\begin{align}
&{\mathcal H}^{(M)}{\psi}^{(M)}(x)=\mathcal{E}{\psi}^{(M)}(x),\qquad
{\mathcal H}^{(M)}{\varphi}^{(M)}_j(x)=
\tilde{\mathcal{E}}_j{\varphi}^{(M)}_j(x),\quad 
j=1,\ldots,M,\\
&
{\mathcal H}^{(M)}=-\frac{d^2}{dx^2}+{U}^{(M)}(x),\quad
{U}^{(M)}(x)\eqdef 
U(x)-2\frac{d^2\log|\text{W}[\varphi_1,\ldots,\varphi_M](x)|}{dx^2}.
\label{Mwron}
\end{align}
Here $\breve{\varphi}_j$ in {\rm \eqref{varphiM}} means 
that $\varphi_j(x)$ is excluded from the Wronskian. 
\end{thm}
In order to apply Darboux transformations in 1-d QM, 
various restrictions must be imposed
on seed solutions so that the deformed potential is 
non-singular in the physical region.
Special choices of seed solutions as eigenfunctions $\varphi_j=\phi_{j-1}$, 
and $\varphi_j=\phi_{d_j}$, $j=1,\ldots,M$  
correspond to Crum \cite{crum} and Krein-Adler \cite{krein,adler}
transformations, respectively.

\subsection{Explicit Examples}
\label{sec:Exam}

Here are three {\em elementary examples of  
exactly solvable potentials}.
The {\em prepotential} $w(x)$ determines the potential $U(x)$
of the Hamiltonian as shown in \eqref{oUdef}:
\begin{align}
  \text{H}:\quad&w(x)=-\tfrac12x^2,\quad \phi_0(x)=e^{-x^2/2},
  \hspace{26mm}-\infty<x<\infty,\n
  &\text{harmonic oscillator:}\quad U(x)=x^2-1,\hspace{25mm}\eta(x)=x,
  \label{ex1}\\
  \text{L}:\quad&w(x)=-\tfrac12x^2+g\log x,\quad \phi_0(x;g)=e^{-x^2/2}x^g,
  \qquad g>\tfrac12,\ \ 0<x<\infty,\n
  &\text{radial oscillator:}\quad 
  U(x)=x^2+\frac{g(g-1)}{x^2}-1-2g,\qquad \eta(x)=x^2,
  \label{ex2}\\
  \text{J}:\quad&w(x)=g\log\sin x+h\log\cos x,\ 
  \phi_0(x;g,h)=(\sin x)^g(\cos x)^h,
  \ g>\tfrac12,\ h>\tfrac12,\ \ 0<x<\tfrac{\pi}{2},\n
  &\text{P\"oschl-Teller:}\quad 
  U(x)=\frac{g(g-1)}{\sin^2x}+\frac{h(h-1)}{\cos^2x}-(g+h)^2,
  \quad\eta(x)=\cos2x.
  \label{ex3}
\end{align}
Their eigenfunctions  have a factorised form:
\begin{equation}
  \phi_n(x)=\phi_0(x)P_n(\eta(x)),\quad \phi_0(x)=e^{w(x)},
  \label{facteig}
\end{equation}
in which $P_n(\eta(x))$ is a degree $n$ (except for those discussed in
\S\ref{sec:Exce}) polynomial in the {\em sinusoidal coordinate} $\eta(x)$.
For the above examples, 
they are the three {\em classical orthogonal polynomials};
the Hermite (H), the Laguerre (L) and the Jacobi (J) polynomials 
(for notation, see Appendix):
\begin{align}
  \text{H}:\quad&
  P_n(\eta(x))=H_n(x)\eqdef(2x)^n
  {}_2F_0\Bigl(\genfrac{}{}{0pt}{}{-\frac{n}{2},\,-\frac{n-1}{2}}
  {-}\Bigm|-\frac{1}{x^2}\Bigr),
  \label{defH}\\
  \text{L}:\quad&
  P_n(\eta(x))=L_n^{(g-\frac12)}(x^2)
  \eqdef\frac{(g+\frac12)_n}{n!}
  {}_1F_1\Bigl(\genfrac{}{}{0pt}{}{-n}{g+\frac12}\Bigm|x^2\Bigr),
  \label{defL}\\
  \text{J}:\quad&
   P_n(\eta(x))=P_n^{(g-\frac12,h-\frac12)}(\cos 2x)
  \eqdef\frac{(g+\frac12)_n}{n!}
  {}_2F_1\Bigl(\genfrac{}{}{0pt}{}{-n,n+g+h}{g+\frac12}\Bigm|
  \frac{1-\cos2x}{2}\Bigr).
  \label{defJ}
\end{align}

The similarity transformed Hamiltonian
$\widetilde{\mathcal{H}}$ in terms of the ground state wavefunction
$\phi_0(x)$
provides the second order equation for $P_n(\eta(x))$
\begin{equation}
 \widetilde{\mathcal{H}}P_n(\eta(x))=\mathcal{E}(n)P_n(\eta(x)), \quad
 \widetilde{\mathcal{H}}\eqdef
  \phi_0(x)^{-1}\circ\mathcal{H}\circ\phi_0(x) 
  =-\frac{d^2}{dx^2}-2\frac{dw(x)}{dx}\frac{d}{dx}.
 \label{polyeq}
\end{equation}
The exact solvability can be rephrased as the  {\em lower triangularity\/} 
of $\widetilde{\mathcal{H}}$
\begin{equation}
  \widetilde{\mathcal{H}}\eta(x)^n
  =\mathcal{E}(n)\eta(x)^n+\text{lower orders in}\ \eta(x),
  \label{lowtri}
\end{equation}
in the special basis
\[
  1,\ \eta(x),\ \eta(x)^2,\ \ldots, \eta(x)^n,\ \ldots,
\]
spanned by the sinusoidal coordinate $\eta(x)$.
This situation is expressed as
\begin{equation}
  \widetilde{\mathcal{H}}\mathcal{V}_n\subseteq\mathcal{V}_n,\quad
  \mathcal{V}_n\eqdef
  \text{Span}\bigl[1,\eta(x),\ldots,\eta(x)^n\bigr].
  \label{Vndef}
\end{equation}

Obviously, the square of the ground state wavefunction 
$\phi_0(x)^2=e^{2w(x)}$ provides
the positive definite orthogonality weight function for the polynomials:
\begin{align}
 & \int_{x_1}^{x_2}\!\!\phi_0(x)^2P_n(\eta(x))P_n(\eta(x))dx=h_n\delta_{nm},
  \label{ortho1}\\
 &\quad  h_n=\left\{\begin{array}{ll}
  2^nn!\sqrt{\pi}&:\text{H}\\[2pt]
  \frac{1}{2\,n!}\,\Gamma(n+g+\tfrac12)&:\text{L}\\[2pt]
  {\displaystyle
  \frac{\Gamma(n+g+\frac12)\Gamma(n+h+\frac12)}
  {2\,n!\,(2n+g+h)\Gamma(n+g+h)}}&:\text{J}
  \end{array}\right..
  \label{hnoQM}
 \end{align}
Let us emphasise that the weight function, 
or $\phi_0(x)$ is determined as a
solution of a first order differential  equation \eqref{Aphi0=0}.

Let us note that $x=0$ for L and $x=0,\pi/{2}$ for J are the
{\em regular singular points} of second order differential equations.
The {\em monodromy} at the regular singular point is determined by the
{\em characteristic exponent} $\rho$:
\begin{equation}
  M_\rho=e^{2\pi i\rho}.
  \label{monodromy}
\end{equation}
The corresponding exponents are expressed simply by the original parameters: 
\begin{equation}
\rho=g, \ 1-g\  \text{for \ L \ and}\quad  \rho=g,\ 1-g\ (x=0),
\quad  \rho=h,\quad 1-h\  (x={\pi}/2) \quad \text{for J}.
\label{monodromy2}
\end{equation}
It is obvious that the radial oscillator \eqref{ex2} 
and the P\"oschl-Teller \eqref{ex3} 
potentials without the constant terms ($-(1+2g)$, $-(g+h)^2$)  are invariant 
under the following {\em discrete symmetry transformations}:
\begin{equation}
\text{L:}\quad g\leftrightarrow 1-g \ ;\quad \text{J:}\quad 
g\leftrightarrow 1-g\ \text{and/or}\ \  h\leftrightarrow 1-h.
\label{dicssymm}
\end{equation}
The same transformations also keep  the above characteristic exponents
\eqref{monodromy2} invariant. Likewise, 
the Hamiltonians of the harmonic \eqref{ex1} and the radial
\eqref{ex2} oscillators (without the constant term) 
{\em change sign under the discrete transformation}
of the coordinate, $x\to i x$:
\begin{align}
 \mathcal{H}_h\to-\mathcal{H}_h,\quad 
 \mathcal{H}_{r}\to-\mathcal{H}_r;\quad
\mathcal{H}_h\eqdef-\partial_x^2+x^2,\quad 
\mathcal{H}_r\eqdef-\partial_x^2+x^2+\frac{g(g-1)}{x^2}.
\label{dicssymm2}
\end{align}

In the next section we will derive these results \eqref{ex1}--\eqref{hnoQM}
based on {\em shape invariance}.

\section{Shape Invariance: Sufficient Condition of Exact Solvability} 
\label{sec:Sha}

Shape invariance \cite{genden} is a sufficient condition for the exact
solvability in the Schr\"{o}dinger picture. Combined with Crum's theorem
\cite{crum}, or the factorisation method \cite{infhull} or the so-called
supersymmetric quantum mechanics \cite{dks,susyqm}, the totality of the
discrete eigenvalues and the corresponding eigenfunctions can be easily
obtained as shown in \eqref{shapeenergy} and \eqref{phin=A..Aphi0}.

\subsection{Energy Spectrum and Rodrigues Formulas} 
\label{sec:Rod}

In many cases the  Hamiltonian contains some parameter(s),
$\bm{\lambda}=(\lambda_1,\lambda_2,\ldots)$.
Here we write the parameter dependence symbolically, 
$\mathcal{H}(\bm{\lambda})$,
$\mathcal{A}(\bm{\lambda})$, $\mathcal{E}(n;\bm{\lambda})$,
$\phi_n(x;\bm{\lambda})$, $P_n(\eta(x);\bm{\lambda})$ etc,
since it is the central issue.
The shape invariance condition with a suitable choice of parameters is
\begin{equation}
  \mathcal{A}(\bm{\lambda})\mathcal{A}(\bm{\lambda})^{\dagger}
  =\mathcal{A}(\bm{\lambda}+\bm{\delta})^{\dagger}
  \mathcal{A}(\bm{\lambda}+\bm{\delta})+\mathcal{E}(1;\bm{\lambda}),
  \label{shapeinv}
\end{equation}
where  $\bm{\delta}$ is the
shift of the parameters. In other words $\mathcal{H}^{[0]}$ and
$\mathcal{H}^{[1]}-\mathcal{E}(1;\bm{\lambda})$ have the same shape, 
only the parameters are
shifted by $\bm{\delta}$.
The $s$-th step Hamiltonian $\mathcal{H}^{[s]}$ in \S\,\ref{sec:Inter}
is $\mathcal{H}^{[s]}=\mathcal{H}(\bm{\lambda}+s\bm{\delta})
+\mathcal{E}(s;\bm{\lambda})$. The energy spectrum and the excited state
wavefunctions are determined by the data of the ground state wavefunction
$\phi_0(x;\bm{\lambda})$ and the energy of the first excited state
$\mathcal{E}(1;\bm{\lambda})$ as follows \cite{dks}:
\begin{align}
  &\mathcal{E}(n;\bm{\lambda})=\sum_{s=0}^{n-1}
  \mathcal{E}(1;\bm{\lambda}^{[s]}),\qquad
  \bm{\lambda}^{[s]}\eqdef\bm{\lambda}+{s}\bm{\delta},
  \label{shapeenergy}\\
  &\phi_n(x;\bm{\lambda})\propto
  \mathcal{A}(\bm{\lambda}^{[0]})^{\dagger}
  \mathcal{A}(\bm{\lambda}^{[1]})^{\dagger}
  \mathcal{A}(\bm{\lambda}^{[2]})^{\dagger}
  \cdots
  \mathcal{A}(\bm{\lambda}^{[n-1]})^{\dagger}
  \phi_0(x;\bm{\lambda}^{[n]}).
  \label{phin=A..Aphi0}
\end{align}
The above formula for the eigenfunctions $\phi_n(x;\bm{\lambda})$ can be
considered as the {\em universal Rodrigues formula\/} for the
 {\em classical orthogonal polynomials}.
For the explicit form of the Rodrigues type formula for each polynomial,
one only has to substitute the explicit forms of the operator
$\mathcal{A}(\bm{\lambda})$ and the ground state wavefunction
$\phi_0(x;\bm{\lambda})$. 

In the case of a finite number of bound states, e.g. the Morse potential,
the eigenvalue has a maximum at a certain level $n$,
$\mathcal{E}(n;\bm{\lambda})$. Beyond that level the formula
\eqref{shapeenergy} ceases to work and the Rodrigues formula
\eqref{phin=A..Aphi0} does not provide  square integrable eigenfunctions,
although $\phi_m$ ($m>n$) continues to satisfy the Schr\"{o}dinger equation
with $\mathcal{E}(m;\bm{\lambda})$.

The above shape invariance condition \eqref{shapeinv} is equivalent to
the following condition:
\begin{align}
 &\bigl(\partial_xw(x;\bm{\lambda})\bigr)^2
  -\partial_x^2w(x;\bm{\lambda})=
  \bigl(\partial_xw(x;\bm{\lambda}+\bm{\delta})\bigr)^2
  +\partial_x^2w(x;\bm{\lambda}+\bm{\delta})+\mathcal{E}(1;\bm{\lambda}).
  \label{shapeinvoQM}
\end{align}

It is straightforward to verify the shape invariance for the three examples
\eqref{ex1}--\eqref{ex3} in \S\ref{sec:Exam} with the following data:
\begin{align}
 \ \text{H}:\quad&
  \bm{\lambda}=\phi\ (\text{null}),\quad\bm{\delta}=\phi,\quad
  \mathcal{A}=\partial_x+x,\qquad\
  \mathcal{E}(n)=2n,
  \label{HEn}\\
  \phantom{\text{oQM}:\ \ }\text{L}:\quad&
  \bm{\lambda}=g,\quad\bm{\delta}=1,\quad
  \mathcal{A}(g)=\partial_x+x-g/x,\quad
\mathcal{E}(n;g)=4n,
\label{LEn}\\
  \phantom{\text{oQM}:\ \ }\text{J}:\quad&
  \bm{\lambda}=(g,h),\quad\bm{\delta}=(1,1),\quad
  \mathcal{A}(g,h)=\partial_x-g\cot x+h\tan x,\\
&\qquad \qquad \qquad \qquad \qquad \qquad \quad
\mathcal{E}(n;g,h)=4n(n+g+h).
\label{JEn}
\end{align}
It should be stressed that the above shape invariant transformation
$\bm{\lambda}\to\bm{\lambda}+\bm{\delta}$,
$\mathcal{H}^{[s]}\to \mathcal{H}^{[s+1]}$ for L and J, that is,
$g\to g+1$, $h\to h+1$, {\em preserves the monodromy} \eqref{monodromy}
at the regular singularities.

The above universal Rodrigues formula \eqref{phin=A..Aphi0} 
for the harmonic oscillator (H) reads
\begin{equation*}
e^{-x^2/2}P_n(\eta)\propto (-\partial_x+x)^n e^{-x^2/2}.
\end{equation*}
By using the relation 
$\partial_x-x=e^{x^2/2}\circ\frac{d}{dx}\circ e^{-x^2/2}$, this gives 
$P_n(\eta)\propto (-1)^ne^{x^2}\left(\frac{d}{dx}\right)^n e^{-x^2}$ 
and the Rodrigues formula for the Hermite
polynomial \eqref{HRod} is obtained. 
The universal Rodrigues formula \eqref{phin=A..Aphi0} for the 
radial  oscillator (L) reads
\begin{equation*}
e^{-x^2/2}x^gP_n(\eta)\propto (-\partial_x+x-\frac{g}{x})\cdots 
(-\partial_x+x-\frac{g+n-1}{x}) e^{-x^2/2}x^{g+n}.
\end{equation*}
By using the relation ($\eta=x^2$)
\begin{equation*}
-\partial_x+x-\frac{g}{x}=-e^{x^2/2}x^{-g}\circ 
\frac{d}{dx}\circ e^{-x^2/2}x^{g}=-2 e^{\eta/2}\eta^{-(g-1)/2}\circ
\frac{d}{d\eta}\circ e^{-\eta/2}\eta^{g/2},
\end{equation*}
the above formula gives
\begin{equation*}
P_n(\eta)\propto (-2)^n e^\eta \eta^{-g+1/2} 
\left(\frac{d}{d\eta}\right)^n \left(e^{-\eta} \eta^{n+g-1/2}\right),
\end{equation*}
and the Rodrigues formula for the Laguerre polynomial \eqref{LRod} is obtained, 
up to an $n$ dependent normalisation constant. 
The verification of the Rodrigues formula \eqref{JRod}
of the Jacobi polynomial  based on \eqref{phin=A..Aphi0} 
is left to readers as an exercise.

\subsection{Forward and Backward Shift Operators} 
\label{sec:FBop}
The shape invariance and the Crum's theorem imply that
$\phi_n(x;\bm{\lambda})$ and $\phi_{n-1}(x;\bm{\lambda}+\bm{\delta})$ are
mapped to each other by the operators $\mathcal{A}(\bm{\lambda})$ and
$\mathcal{A}(\bm{\lambda})^{\dagger}$:
\begin{align}
  \mathcal{A}(\bm{\lambda})\phi_n(x;\bm{\lambda})
  &=f_n(\bm{\lambda})
  \phi_{n-1}\bigl(x;\bm{\lambda}+\bm{\delta}\bigr),
  \label{Aphi=fphi}\\
  \mathcal{A}(\bm{\lambda})^{\dagger}
  \phi_{n-1}\bigl(x;\bm{\lambda}+\bm{\delta}\bigr)
  &=b_{n-1}(\bm{\lambda})\phi_n(x;\bm{\lambda}).
  \label{Adphi=bphi}
\end{align}
Here the constants $f_n(\bm{\lambda})$ 
and $b_{n-1}(\bm{\lambda})$ depend
on the normalisation of $\{\phi_n(x;\bm{\lambda})\}$ 
but their product does not. It gives the energy eigenvalue,
\begin{equation}
  \mathcal{E}(n;\bm{\lambda})=f_n(\bm{\lambda})b_{n-1}(\bm{\lambda}).
\end{equation}
 For our choice of $\phi_n(x;\bm{\lambda})$ and
$P_n(\eta(x))$ \eqref{defH}--\eqref{defJ}, 
the data for $f_n(\bm{\lambda})$
and $b_{n-1}(\bm{\lambda})$ are:
\begin{align}
&f_n(\bm{\lambda})=\left\{
  \begin{array}{ll}
  2n&:\text{H}\\
  -2&:\text{L}\\
  -2(n+g+h)&:\text{J}
  \end{array}\right.\!,
  \quad
  b_{n-1}(\bm{\lambda})=\left\{
  \begin{array}{ll}
  1&:\text{H}\\
  -2n&:\text{L,J}
  \end{array}\right.\!.
  \label{fnbn}
\end{align}
By removing the ground state contributions, 
the {\em forward and backward shift
operators} acting on the polynomial eigenfunctions,
$\mathcal{F}(\bm{\lambda})$ and $\mathcal{B}(\bm{\lambda})$, are introduced:
\begin{align}
  \mathcal{F}(\bm{\lambda})&\eqdef
  \phi_0(x;\bm{\lambda}+\bm{\delta})^{-1}\circ
  \mathcal{A}(\bm{\lambda})\circ\phi_0(x;\bm{\lambda}) =\cF\frac{d}{d\eta},
  \label{Fdef}\\
  \mathcal{B}(\bm{\lambda})&\eqdef
  \phi_0(x;\bm{\lambda})^{-1}\circ
  \mathcal{A}(\bm{\lambda})^{\dagger}
  \circ\phi_0(x;\bm{\lambda}+\bm{\delta})
    \label{Bdef}\\
  &=
  -4\cF^{-1}c_2(\eta)\bigl(\frac{d}{d\eta}
  +\frac{c_1(\eta,\bm{\lambda})}{c_2(\eta)}\bigr),
  \end{align}
where $\cF$, $c_1(\eta,\bm{\lambda})$ and $c_2(\eta)$ are
\begin{equation}
  \cF\eqdef\left\{
  \begin{array}{ll}
  1&:\text{H}\\
  2&\!:\text{L}\\
  -4&\!:\text{J}
  \end{array}\right.\!\!,
  \ \ c_1(\eta,\bm{\lambda})\eqdef\left\{
  \begin{array}{ll}
  -\frac12&\!:\text{H}\\
  g+\tfrac12-\eta&\!:\text{L}\\
  h-g-(g+h+1)\eta&\!:\text{J}
  \end{array}\right.\!\!,
  \ \ c_2(\eta)\eqdef\left\{
  \begin{array}{ll}
  \frac14&\!:\text{H}\\
  \eta&\!:\text{L}\\
  1-\eta^2&\!:\text{J}
  \end{array}\right.\!\!.\!\!
  \label{cF,c1,c2}
\end{equation}
Then the above relations \eqref{Aphi=fphi}--\eqref{Adphi=bphi} become
\begin{align}
 \cF\frac{d}{d\eta}{P}_n(\eta;\bm{\lambda})
  &=f_n(\bm{\lambda}){P}_{n-1}(\eta;\bm{\lambda}+\bm{\delta}),
  \label{forwardrel}\\
  \mathcal{B}(\bm{\lambda}){P}_{n-1}(\eta;\bm{\lambda}+\bm{\delta})
  &=b_{n-1}(\bm{\lambda}){P}_n(\eta;\bm{\lambda}).
  \label{backwardrel}
\end{align}
Corresponding to \eqref{factHam}, 
the forward and backward shift operators
give  the similarity transformed Hamiltonian \eqref{polyeq}:
\begin{equation}
  \widetilde{\mathcal{H}}(\bm{\lambda})
  =\mathcal{B}(\bm{\lambda})\mathcal{F}(\bm{\lambda})=
  -4\bigl(c_2(\eta)\frac{d^2}{d\eta^2}
  +{c_1(\eta,\bm{\lambda})}\frac{d}{d\eta}\bigr),
  \label{htilbf}
\end{equation}
which provides the well known forms of the differential equations 
for the classical orthogonal polynomials.
It is trivial to verify the lower triangularity of 
$\widetilde{\mathcal H}$ \eqref{lowtri} 
for the three examples \eqref{defH}--\eqref{defJ} based on \eqref{htilbf}.

Let us conclude this section by  remarks on the effects of Crum \S\ref{sec:Inter} and 
Krein-Adler \S\ref{sec:ModCrum} transformations of shape invariant systems.
By deleting $M$ successive ground states in terms of Crum transformations,
one arrives at the same theory with the shifted parameters, 
$\bm{\lambda}\to \bm{\lambda}+M\bm{\delta}$. Thus an  essentially  new theory is not
created by Crum transformations on shape invariant systems.
On the other hand, by applying Krein-Adler transformations on a shape invariant system,
one can create infinitely new solvable systems as explained in \S\ref{sec:Boch}.
\section{Solvability in the Heisenberg Picture}
\label{sec:Hei}

\subsection{Closure Relations}
\label{sec:close}

As is well known the Heisenberg operator formulation is central to
quantum field theory. The creation/annihilation operators of the
harmonic oscillators are the cornerstones of modern quantum physics.
However, until recently, it had been generally conceived that the
Heisenberg operator solutions are intractable. Here we show that most of
the shape invariant QM Hamiltonian systems are exactly solvable in
the Heisenberg picture, too \cite{os7,os8}. To be more precise, the
Heisenberg operator of the sinusoidal coordinate $\eta(x)$
\begin{equation}
  e^{it\mathcal{H}}\eta(x)e^{-it\mathcal{H}}
  \label{etaHei}
\end{equation}
can be evaluated in a closed form. It is well known that any orthogonal
polynomials starting at degree 0 satisfy the {\em three term recurrence relations}
\cite{askey,ismail}
\begin{equation}
  \eta P_n(\eta)=A_nP_{n+1}(\eta)+B_nP_n(\eta)+C_nP_{n-1}(\eta)\ \ (n\geq 0),
  \label{threeterm}
\end{equation}
with $P_{0}(\eta)=\text{constant}$,  $P_{-1}(\eta)=0$. 
Here the coefficients $A_n$, $B_n$ and $C_n$ 
depend on the normalisation of $\{P_n(\eta)\}$.
They are real and $A_{n-1}C_n>0$ ($n\geq 1$). Conversely all the polynomials
starting with degree 0 and satisfying the above three term recurrence
relations are orthogonal (Favard's theorem \cite{Chihara}).

For the factorised quantum mechanical eigenfunctions \eqref{facteig},
these relations mean
\begin{equation}
  \eta(x)\phi_n(x)=A_n\phi_{n+1}(x)+B_n\phi_n(x)+C_n\phi_{n-1}(x)
  \ \ (n\geq 0),\quad \phi_{-1}(x)=0.
  \label{threetermphi}
\end{equation}
In other words, the operator $\eta(x)$ acts like a creation operator
which sends the eigenstate $n$ to $n+1$ as well as like an annihilation
operator, which maps an eigenstate $n$ to $n-1$.
This fact combined with the well known result that the annihilation/creation
operators of the harmonic oscillator are the positive/negative frequency
parts of the Heisenberg operator solution for the coordinate $x$ is the
starting point of this subsection.
As will be shown below the {\em sinusoidal coordinate} $\eta(x)$ undergoes
sinusoidal motion \eqref{quantsol}, whose frequencies depend on the energy.
Thus it is not harmonic in general.
To the best of our knowledge, the sinusoidal coordinate was first introduced
in a rather broad sense for general (not necessarily solvable) potentials as
a useful means for coherent state research by Nieto and Simmons \cite{nieto}.

A sufficient condition for the closed form expression of the Heisenberg
operator \eqref{etaHei} is the {\em closure relation\/}
\begin{equation}
  [\mathcal{H},[\mathcal{H},\eta(x)]\,]
  =\eta(x)\,R_0(\mathcal{H})+[\mathcal{H},\eta(x)]\,R_1(\mathcal{H})
  +R_{-1}(\mathcal{H}).
  \label{closurerel}
\end{equation}
Here the coefficients $R_i(y)$ are polynomials in $y$. It is easy to see
that the cubic commutator $[\mathcal{H},[\mathcal{H},[\mathcal{H},\eta(x)]]]
\equiv(\text{ad}\,\mathcal{H})^3\eta(x)$ is reduced to $\eta(x)$ and
$[\mathcal{H},\eta(x)]$ with $\mathcal{H}$ depending coefficients:
\begin{align*}
  (\text{ad}\,\mathcal{H})^3\eta(x)&=
  [\mathcal{H},\eta(x)]R_0(\mathcal{H})
  +[\mathcal{H},[\mathcal{H},\eta(x)]]\,R_1(\mathcal{H})\n
  &=\eta(x)\,R_0(\mathcal{H})R_1(\mathcal{H})
  +[\mathcal{H},\eta(x)]\,\bigl(R_1(\mathcal{H})^2+R_0(\mathcal{H})\bigr)
  +R_{-1}(\mathcal{H})R_{1}(\mathcal{H}),
\end{align*}
in which the definition $(\text{ad}\,\mathcal{H})X\eqdef[\mathcal{H},X]$
is used. In this notation the above closure relation \eqref{closurerel} reads
\begin{equation}
(\text{ad}\,\mathcal{H})^2\eta(x)
  =\eta(x)\,R_0(\mathcal{H})+(\text{ad}\,\mathcal{H})\eta(x)\,R_1(\mathcal{H})
  +R_{-1}(\mathcal{H}),
  \label{closurerel2}
\end{equation}
which can be understood as the {\em Cayley-Hamilton equation} for the operator 
$\text{ad}\,\mathcal{H}$ acting on $\eta(x)$.
It is trivial to see that all the higher commutators
$(\text{ad}\,\mathcal{H})^n\eta(x)$ can also be reduced to $\eta(x)$ and
$[\mathcal{H},\eta(x)]$ with $\mathcal{H}$ depending coefficients.
The second order closure \eqref{closurerel} simply reflects the
Schr\"{o}dinger equation, which is a second order differential 
equation. Thus we arrive at
\begin{align}
  e^{it\mathcal{H}}\eta(x)e^{-it\mathcal{H}}
  &=\sum_{n=0}^\infty\frac{(it)^n}{n!}(\text{ad}\,\mathcal{H})^n\eta(x)\n
  &=[\mathcal{H},\eta(x)]
  \frac{e^{i\alpha_+(\mathcal{H})t}-e^{i\alpha_-(\mathcal{H})t}}
  {\alpha_+(\mathcal{H})-\alpha_-(\mathcal{H})}
  -R_{-1}(\mathcal{H})R_{0}(\mathcal{H})^{-1}\n
  &\quad
  +\bigl(\eta(x)+R_{-1}(\mathcal{H})R_0(\mathcal{H})^{-1}\bigr)
  \frac{-\alpha_-(\mathcal{H})e^{i\alpha_+(\mathcal{H})t}
  +\alpha_+(\mathcal{H})e^{i\alpha_-(\mathcal{H})t}}
  {\alpha_+(\mathcal{H})-\alpha_-(\mathcal{H})}.
  \label{quantsol}
\end{align}
This simply means that $\eta(x)$ oscillates sinusoidally with two
energy-dependent ``frequencies'' $\alpha_\pm(\mathcal{H})$ given by
\begin{gather}
  \alpha_\pm(\mathcal{H})\eqdef\tfrac12\bigl(R_1(\mathcal{H})\pm
  \sqrt{R_1(\mathcal{H})^2+4R_0(\mathcal{H})}\,\bigr),
  \label{alpmdef}\\
  \alpha_+(\mathcal{H})+\alpha_-(\mathcal{H})=R_1(\mathcal{H}),
  \quad
  \alpha_+(\mathcal{H})\alpha_-(\mathcal{H})=-R_0(\mathcal{H}).
  \label{freqpm}
\end{gather}
The energy spectrum is determined by the over-determined recursion relations
\begin{equation}
\mathcal{E}(n+1)=\mathcal{E}(n)+\alpha_+(\mathcal{E}(n)),\quad
\mathcal{E}(n-1)=\mathcal{E}(n)+\alpha_-(\mathcal{E}(n)),\quad 
\mathcal{E}(0)=0. 
\label{HeiEn}
\end{equation}
It should be stressed that for the known spectra $\{\mathcal{E}(n)\}$
determined by the shape invariance, the quantity inside the square root
in the definition of $\alpha_\pm(\mathcal{H})$ \eqref{alpmdef} for each $n$:
\[
  R_1(\mathcal{E}(n))^2+4R_0(\mathcal{E}(n))
\]
becomes a complete square and the the above two conditions are consistent.
For 1-d QM, the Hamiltonians and the sinusoidal coordinates satisfying the closure
relation \eqref{closurerel} are classified and then the eigenfunctions have
the factorised form \eqref{facteig} \cite{os7}.
For the three elementary examples \eqref{ex1}--\eqref{ex3} 
given in \S\ref{sec:Exam}, the data are:
\begin{align}
  \text{H}:\quad&R_1(y)=0,\quad R_0(y)=4,\quad R_{-1}(y)=0,\\
  & A_n=1/2,\quad B_n=0,\quad C_n=1,\\
  \text{L}:\quad&R_1(y)=0,\quad R_0(y)=16,\quad R_{-1}(y)=-8(y+2g+1),\\
& A_n=-(n+1),\quad B_n=(2n+g+1/2),\quad C_n=-(n+g-1/2),\\
  \text{J}:\quad&R_1(y)=8,\quad R_0(y)=16\bigl(y+(g+h)^2-1\bigr),\quad
  R_{-1}(y)=16(g-h)(g+h-1),\\[2pt]
& A_n= \frac{2(n+1)(n+g+h)}
 {(2n+g+h)(2n+g+h+1)},\quad B_n=\frac{(h-g)(g+h-1)}
 {(2n+g+h-1)(2n+g+h+1)},\n[2pt]
& C_n=\frac{2(n+g-1/2)(n+h-1/2)}
 {(2n+g+h-1)(2n+g+h)}.
\end{align}
It is straight forward to verify the recursion relations \eqref{HeiEn} 
for the eigenvalue formulas $\mathcal{E}(n)$,
\eqref{HEn}--\eqref{JEn} for the three elementary examples.

For 1-d QM, the necessary and sufficient conditions for the existence of the
sinusoidal coordinate satisfying the closure relation \eqref{closurerel}
are analysed in Appendix A of \cite{os7}. It was shown that such systems
constitute a sub-group of the shape invariant 1-d QM. We also mention that
exact Heisenberg operator solutions for independent sinusoidal coordinates
as many as the degree of freedom were derived for the Calogero systems
based on any root system \cite{os9}. These are novel examples of
infinitely many multi-particle Heisenberg operator solutions.

\subsection{Annihilation and Creation Operators}
\label{sec:an-cr}

The {\em annihilation} and {\em creation} operators $a^{(\pm)}$ are
extracted from this exact Heisenberg operator solution:
\begin{align}
  &e^{it\mathcal{H}}\eta(x)e^{-it\mathcal{H}}
  =a^{(+)}e^{i\alpha_+(\mathcal{H})t}+a^{(-)}e^{i\alpha_-(\mathcal{H})t}
  -R_{-1}(\mathcal{H})R_0(\mathcal{H})^{-1},
  \label{apm}\\
  &a^{(\pm)}\eqdef\pm\Bigl([\mathcal{H},\eta(x)]-\bigl(\eta(x)
  +R_{-1}(\mathcal{H})R_0(\mathcal{H})^{-1}\bigr)\alpha_{\mp}(\mathcal{H})
  \Bigr)
  \bigl(\alpha_+(\mathcal{H})-\alpha_-(\mathcal{H})\bigr)^{-1}\n
  &\phantom{a^{(\pm)}}=
  \pm\bigl(\alpha_+(\mathcal{H})-\alpha_-(\mathcal{H})\bigr)^{-1}
  \Bigl([\mathcal{H},\eta(x)]+\alpha_{\pm}(\mathcal{H})\bigl(\eta(x)
  +R_{-1}(\mathcal{H})R_0(\mathcal{H})^{-1}\bigr)\Bigr),
  \label{apmdefs}\\
  & a^{(+)\,\dagger}=a^{(-)},\quad
  a^{(+)}\phi_n(x)=A_n\phi_{n+1}(x),\quad
  a^{(-)}\phi_n(x)=C_n\phi_{n-1}(x).
  \label{apmphi}
\end{align}
It should be stressed that the annihilation  and the creation operators
are hermitian conjugate of each other and they act on the eigenstate
\eqref{apmphi}.
The excited state wavefunctions $\{\phi_n(x)\}$ are obtained by the
successive action of the creation operator $a^{(+)}$ on the ground state
wavefunction $\phi_0(x)$.
This is the exact solvability in the Heisenberg picture.

\subsection{Dynamical Symmetry Algebras and Coherent States}
\label{sec:cohe}

Simple commutation relations
\begin{equation}
  [\mathcal{H},a^{(\pm)}]=a^{(\pm)}\alpha_{\pm}(\mathcal{H}),
  \label{[H,apm]}
\end{equation}
follow from \eqref{apmdefs} and \eqref{closurerel}.
Commutation relations of $a^{(\pm)}$ are expressed in terms of the
coefficients of the three term recurrence relations by \eqref{apmphi}:
\begin{align}
  &a^{(-)}a^{(+)}\phi_n=A_nC_{n+1}\phi_n,\quad
  a^{(+)}a^{(-)}\phi_n=C_nA_{n-1}\phi_n,\n
  &\Rightarrow
  \ [a^{(-)},a^{(+)}]\phi_n=(A_nC_{n+1}-A_{n-1}C_n)\phi_n.
  \label{[a-,a+]}
\end{align}
These relations simply mean the operator relations
\begin{gather}
  a^{(-)}a^{(+)}=f(\mathcal{H}),\quad
  a^{(+)}a^{(-)}=g(\mathcal{H}),
  \label{apamg}
\end{gather}
in which $f$ and $g$ are analytic functions of $\mathcal{H}$ explicitly
given for each example. In other words, $\mathcal{H}$ and $a^{(\pm)}$ form
a {\em dynamical symmetry algebra}, 
which is also called a {\em quasi-linear algebra} \cite{vinzhed}.
It should be stressed that the situation is quite different from those
of the wide variety of proposed annihilation/creation operators for various
quantum systems \cite{coherents}, most of which were introduced within
the framework of `algebraic theory of coherent states,' without exact
solvability. In all such cases there is no guarantee for symmetry
relations like \eqref{apamg}.
The explicit form of the annihilation operator \eqref{apmdefs} allows us to
define the {\em coherent state} as its eigenvector,
$a^{(-)}\psi(\alpha,x)=\alpha\psi(\alpha,x)$, $\alpha\in\mathbb{C}$.
See \cite{os7} for various examples of coherent states.

\subsection{Bochner's Theorem}
\label{sec:Boch}

In 1884 \cite{routh},  Routh showed that polynomials satisfying the three
term recurrence relations and a second order differential equation were
one of the {\em classical} polynomials, the Hermite, Laguerre, Jacobi
and Bessel. 
Later in 1929 \cite{bochner}, Bochner classified all polynomial solutions to
second order Sturm-Liouville operators with polynomial coefficients and
arrived at the same conclusions.
See \S\,20.1 of \cite{ismail} for more details.
This was a kind of No-Go theorem in 1-d QM, since it declared
that no essentially new exactly solvable 1-d QM could be achieved as the
solutions of the ordinary Schr\"{o}dinger equation under the assumption of the factorised
eigenfunctions \eqref{facteig} with $n$ specifying the degree of the polynomial.
Avoiding Bochner's theorem was one of the strongest motivations
for the introduction of the Askey scheme
of hypergeometric orthogonal polynomials 
and their $q$-analogues \cite{askey, ismail, koekswart}.
The {\em discrete quantum mechanics} \cite{os24} 
was created as a quantum mechanical reformulation
of these Askey scheme of classical orthogonal polynomials.
In the discrete quantum mechanics Schr\"{o}dinger equations are difference equations 
instead of differential, 
and the constraints by Bochner's theorem do not apply.
Many exactly solvable examples of discrete quantum mechanics have been constructed
\cite{os12,os13}.

\bigskip
One simple way to avoid Bochner's theorem in 1-d QM 
is to apply Krein-Adler transformations 
on the three examples of classical orthogonal polynomials 
H, L and J, \eqref{defH}--\eqref{defJ}
introduced in \S\ref{sec:Exam}. For 
$\mathcal{D}\eqdef\{d_1,d_2,\ldots,d_{M}\}\subset\mathbb{Z}_{\ge0}^{M}$, 
$\mathcal{D}\neq\{0,1,\ldots,M-1\}$,
the resulting eigenfunction \eqref{adlphin} is \cite{adler}:
\begin{align}
\bar{\phi}_{{\mathcal D};n}(x)&\eqdef
  \frac{\text{W}\,[\phi_{d_1},\phi_{d_2},\ldots,\phi_{d_{M}},\phi_n](x)}
  {\text{W}\,[\phi_{d_1},\phi_{d_2},\ldots,\phi_{d_{M}}](x)}=
  \psi_{\mathcal D}(x)\frac{\mathcal{P}_{\mathcal{D};n}(\eta)}{\Xi_{\mathcal D}(\eta)},
   \quad \eta\equiv\eta(x), \\
\psi_{\mathcal D}(x)&\eqdef 
\phi_0(x)\left(\eta'(x)\right)^{M+1},\qquad \eta'(x)\equiv d\eta(x)/dx,\\
\Xi_{\mathcal D}(\eta)&\eqdef\text{W}\,[P_{d_1},P_{d_2},\ldots,P_{d_{M}}](\eta),\quad
\mathcal{P}_{\mathcal{D};n}(\eta)\eqdef 
\text{W}\,[P_{d_1},P_{d_2},\ldots,P_{d_{M}},P_n](\eta),
\end{align}
in which a formula 
\begin{equation}
\text{W}\,[f_1(\eta(x)),f_2(\eta(x)),\ldots,f_n(\eta(x))](x)=
\left(\eta'(x)\right)^{n(n+1)/2}
\text{W}\,[f_1,f_2,\ldots,f_n](\eta)
\end{equation}
is used. 
Here $\Xi_{\mathcal D}(\eta)$ and $\mathcal{P}_{\mathcal{D};n}(\eta)$ 
are polynomials in $\eta$ of degree $\ell_{D}$ and $\ell_{\mathcal D}+n-M$,
 with $\ell_{\mathcal D}\eqdef\sum_{j=1}^M d_j-M(M-1)/2\ge M$.
Thus the eigenfunctions of the deformed system 
provide orthogonal polynomials over $(x_1,x_2)$, 
$\{\mathcal{P}_{\mathcal{D};n}(\eta)\}$, $n\in\mathbb{Z}_{\ge0}\backslash\mathcal{D}$, 
satisfying second order differential equations. 
The weight function is 
$W_{\mathcal D}(x)\eqdef\psi_{\mathcal D}^2(x)/\Xi_{\mathcal D}^2(\eta)$.
These polynomials, however, have $M$ `{\em holes}' in the degree at $n=d_j$, 
$j=1,\ldots,M$ and the lowest degree
is $\ell_{\mathcal D}+\mu-M\ge0$, in which $\mu$ is defined in \eqref{defmu}.

The simplest example of $\mathcal{D}=\{1,2\}$ 
for the harmonic oscillator (H) potential, 
$W_{\mathcal D}(x)\propto{e^{-x^2}}/{(1+2x^2)^2}$ was derived by 
Dubov, Eleonski\u{i} and Klagin \cite{dubov} based on a pseudo virtual state 
$\tilde{\phi}_2(x)$ \eqref{pvsH} 
deletion  a few years before Adler \cite{adler}.
This is the simplest example of the 
{\em duality between the pseudo virtual states and eigenstates} 
demonstrated in \S\ref{sec:dual}.

\bigskip
In this context, it is clear that orthogonal polynomials avoiding 
Bochner's constraints should start at degree 
$\ell\ge1$, if they do not have `holes' in the degrees.
Thus they do not satisfy the three term recurrence relations \eqref{threeterm}.
The pursuit for such new orthogonal polynomials led 
G\'{o}mez-Ullate, Kamran and Milson \cite{gomez}
in 2008 to the discovery of {\em exceptional orthogonal polynomials},
called $X_1$ Laguerre and Jacobi polynomials, which start at degree 1.
Almost immediately, Quesne proposed quantum mechanical reformulation of the 
$X_1$ Laguerre and Jacobi polynomials as the main parts of the eigenfunctions of 
shape invariant systems \cite{quesne}. 
Odake and Sasaki constructed $X_\ell$ Laguerre and Jacobi polynomials,
which start at degree $\ell$,  for all positive integers $\ell$ \cite{os16}. 
Then Quesne \cite{quesne2} discovered the second type of exceptional Laguerre and Jacobi 
polynomials at $\ell=2$. At $\ell=1$ the first and the second type are identical.
These two types are called $X^{\I\, (\II)}_\ell$-L(J) polynomials 
for short and they are generated by the
discrete symmetry transformations \eqref{dicssymm}--\eqref{dicssymm2} 
of the radial oscillator and the P\"oschl-Teller potentials.
Odake and Sasaki \cite{os19} constructed the $X_\ell^{\II}$-Laguerre and 
Jacobi polynomials for all positive integers $\ell$.
Soon this exciting hot topic attracted many authors 
and the rich structures of the subjects
\cite{junkroy}--\cite{refl}, including for example, 
identities satisfied by orthogonal polynomials 
\cite{os18,os29}, Fuchsian differential equations aspects 
\cite{hos, gomez4, st3, hst, heun}, 
Darboux transformations \cite{gomez2, stz, gomez3,os25}, alternatives of the
three term recurrence relations \cite{stz,rec}, etc
were revealed.

\section{New Orthogonal Polynomials}
\label{sec:Exce}

In this section we present  infinitely many new orthogonal polynomials satisfying
second order differential  equations, not following the historical developments,
but adhering to the logical structure. 
The main focus is the {\em multi-indexed}
Laguerre and Jacobi polynomials, 
which include the exceptional orthogonal polynomials as the 
simplest one-indexed cases.
The multiple Darboux transformations with polynomial  type seed solutions, 
which are obtained 
from the eigenfunctions through discrete symmetry transformations, 
{\em the virtual state wave functions}, play the central role.
Needless to say, one can apply the Krein-Adler transformations
to multi-indexed orthogonal polynomials to generate the ones 
with the `holes' in the degree.

\subsection{Polynomial Type Seed Solutions}
\label{sec:polyseed}

For rational extensions of solvable potentials, 
we need {\em polynomial type seed solutions},
which are factorised into a prefactor times a polynomial as \eqref{facteig}.
We will introduce three kinds of polynomial type seed solutions, called 
{\em virtual state wavefunctions\/}, 
{\em pseudo virtual state wavefunctions\/} and
{\em overshoot eigenfunctions\/}.
 We will discuss overshoot eigenfunctions in connection with
the deformations of exactly solvable scattering problems in \S\ref{sec:Scat}.

The seed solutions 
$\{\varphi_j(x),\tilde{\mathcal{E}}_j\}$ ($j=1,2,\ldots,M$)
satisfying the following conditions are called
{\em virtual state wavefunctions\/}:
\begin{enumerate}
\item No zeros in $x_1<x<x_2$, {\em i.e.} $\varphi_j(x)>0$ or
$\varphi_j(x)<0$ in $x_1<x<x_2$.
\item Negative energy, $\tilde{\mathcal{E}}_j<0$.
\item  $\varphi_j(x)$ is also a polynomial type solution.
\item Square non-integrability, $(\varphi_j,\varphi_j)=\infty$.
\item Reciprocal square non-integrability,
$(\varphi_j^{-1},\varphi_j^{-1})=\infty$.
\end{enumerate}
These conditions are not totally independent.
The negative energy condition is necessary for the positivity of the norm
as seen from the norm formula \eqref{modnorm}, since a similar formula is
valid for the virtual state wavefunction cases when $\mathcal{E}_{d_j}$
is replaced by $\tilde{\mathcal{E}}_j$.

When the first condition is dropped and the reciprocal is required to be
locally square integrable 
{\em at both boundaries\/},  see \eqref{type3},
such seed solutions are called {\em pseudo virtual state wavefunctions}.
When the system is extended in terms of a pseudo virtual state wavefunction
$\varphi_j(x)$, the new Hamiltonian $\mathcal{H}^{(1)}$  has an extra
{\em eigenstate} $\varphi_j^{-1}(x)$ with the eigenvalue
$\tilde{\mathcal{E}}_j$, {\em if the new potential is non-singular}.
The extra state is below the original ground state and $\mathcal{H}^{(1)}$
is no longer iso-spectral with $\mathcal{H}$. This is a consequence of
\eqref{newschr}. Its non-singularity is not guaranteed, either.
When extended in terms of $M$ pseudo virtual state wavefunctions
$\{\varphi_j(x),\tilde{\mathcal{E}}_j\}$ ($j=1,2,\ldots,M$), the resulting
Hamiltonian $\mathcal{H}^{(M)}$ has $M$ additional {\em eigenstates}
$\breve{\varphi}_j^{(M)}(x)$ \eqref{varphiM}, {\em if the potential
$U^{(M)}(x)$ is non-singular}. They are all below the original ground state.

Since $\varphi_j(x)$ is finite in $x_1<x<x_2$, the non-square integrability
can only be caused by the boundaries. Thus the virtual state wavefunctions
belong to either of the following type $\I$ and $\II$:
\begin{alignat}{3}
  \text{Type $\I$ virtual}: &&
  &\int_{x_1}^{x_1+\epsilon}\!\!\!dx\,\varphi_j(x)^2<\infty,\quad
  &&\int_{x_2-\epsilon}^{x_2}\!\!\!dx\,\varphi_j(x)^2=\infty,\n
  &\ \ \text{\&}\ &&\int_{x_1}^{x_1+\epsilon}\!\!\!dx\,\varphi_j(x)^{-2}=\infty,
  \quad
  &&\int_{x_2-\epsilon}^{x_2}\!\!\!dx\,\varphi_j(x)^{-2}<\infty,
  \label{type1}\\
  \text{Type $\II$ virtual}:&&
  &\int_{x_1}^{x_1+\epsilon}\!\!\!dx\,\varphi_j(x)^2=\infty,\quad
  &&\int_{x_2-\epsilon}^{x_2}\!\!\!dx\,\varphi_j(x)^2<\infty,\n
  &\ \ \text{\&}\ &&\int_{x_1}^{x_1+\epsilon}\!\!\!dx\,\varphi_j(x)^{-2}<\infty,
  \quad
  &&\int_{x_2-\epsilon}^{x_2}\!\!\!dx\,\varphi_j(x)^{-2}=\infty,
  \label{type2}\\
  \text{pseudo virtual}:&&
  &\int_{x_1}^{x_1+\epsilon}\!\!\!dx\,\varphi_j(x)^2=\infty\ \ \text{or}
  &&\int_{x_2-\epsilon}^{x_2}\!\!\!dx\,\varphi_j(x)^2=\infty,\n
  &\ \ \text{\&}\ &&\int_{x_1}^{x_1+\epsilon}\!\!\!dx\,\varphi_j(x)^{-2}<\infty,
  \quad
  &&\int_{x_2-\epsilon}^{x_2}\!\!\!dx\,\varphi_j(x)^{-2}<\infty.
  \label{type3}
\end{alignat}
An appropriate modification is needed when $x_2=+\infty$ and/or $x_1=-\infty$.

\subsubsection{Virtual state wavefunctions for L and J}
\label{virtLJ}
Here we present the explicit forms of the virtual state wave functions for the
radial oscillator (L) \eqref{ex2} and the P\"oschl-Teller (J) potentials \eqref{ex3}.
They are obtained from their eigenfunctions \eqref{facteig}, \eqref{defL}--\eqref{defJ}
by the discrete symmetry transformations \eqref{dicssymm} and \eqref{dicssymm2} for L1:
The virtual states wavefunctions for L are:
\begin{align}
  \text{L1}:&\quad
  \tilde{\phi}_\text{v}^\ai(x)\eqdef
  e^{\frac12x^2}x^g\xi_\text{v}^\ai(\eta(x);g),\quad
  \xi_\text{v}^\ai(\eta;g)\eqdef P_\text{v}(-\eta;g),\n
  &\quad
  \tilde{\mathcal{E}}_\text{v}^\ai\eqdef-4(g+\text{v}+\tfrac12),\quad
  \text{v}\in\mathbb{Z}_{\ge0},\quad\tilde{\bm{\delta}}^{\ai}\eqdef-1,
  \label{vsL1}\\
  \text{L2}:&\quad
  \tilde{\phi}_\text{v}^\ait(x)\eqdef
  e^{-\frac12x^2}x^{1-g}\xi_\text{v}^\ait(\eta(x);g),\quad
  \xi_\text{v}^\ait(\eta;g)\eqdef P_\text{v}(\eta;1-g),\n
  &\quad
  \tilde{\mathcal{E}}_\text{v}^\ait\eqdef-4(g-\text{v}-\tfrac12),\quad
  \text{v}=0,1,\ldots,[g-\tfrac12]',\quad
  \tilde{\bm{\delta}}^{\ait}\eqdef 1,
  \label{vsL2}
\end{align}
in which $[a]'$ denotes the greatest integer less than $a$ and
$\tilde{\bm{\delta}}^{\ai,\ait}$ will be used later.
For no-nodeness of $\xi_{\text{v}}^{\ai,\ait}$, see (2.39) of \cite{os18}.
The virtual state wavefunctions for J are:
\begin{align}
  \text{J1}:&\quad
  \tilde{\phi}_\text{v}^\ai(x)\eqdef(\sin x)^g(\cos x)^{1-h}
  \xi_\text{v}^\ai(\eta(x);g,h),\quad
  \xi_\text{v}^\ai(\eta;g,h)\eqdef P_\text{v}(\eta;g,1-h),\n
  &\quad
  \tilde{\mathcal{E}}_\text{v}^\ai\eqdef-4(g+\text{v}+\tfrac12)
  (h-\text{v}-\tfrac12),\quad
  \text{v}=0,1,\ldots,[h-\tfrac12]',\quad
  \tilde{\bm{\delta}}^{\ai}\eqdef(-1,1),
  \label{vsJ1}\\
  \text{J2}:&\quad
  \tilde{\phi}_\text{v}^\ait(x)\eqdef(\sin x)^{1-g}(\cos x)^h
  \xi_\text{v}^\ait(\eta(x);g,h),\quad
  \xi_\text{v}^\ait(\eta;g,h)\eqdef P_\text{v}(\eta;1-g,h),\n
  &\quad
  \tilde{\mathcal{E}}_\text{v}^\ait\eqdef-4(g-\text{v}-\tfrac12)
  (h+\text{v}+\tfrac12),\quad
  \text{v}=0,1,\ldots,[g-\tfrac12]',\quad
  \tilde{\bm{\delta}}^{\ait}\eqdef(1,-1).
  \label{vsJ2}
\end{align}
The larger the parameter $g$ and/or $h$ become, the more the virtual states
are `created'. 
The L1 system is obtained from the J1 in the limit $h\to\infty$.
This explains the infinitely many virtual states of L1.
Let us denote by $\mathcal{V}^{\ai,\ait}$ the index sets of
the virtual states of type I and II for L and J.
Due to the parity property of the Jacobi polynomial
$P_n^{(\alpha,\beta)}(-x)=(-1)^nP_n^{(\beta,\alpha)}(x)$, the two virtual
state polynomials $\xi_\text{v}^\ai$ and $\xi_\text{v}^\ait$ for J are
related by
$\xi_\text{v}^\ait(-\eta;g,h)=(-1)^\text{v}\xi_\text{v}^\ai(\eta;h,g)$.
For no-nodeness of $\xi_{\text{v}}^{\ai,\ait}$, see (2.40) of \cite{os18}
and (3.2) of \cite{os21}.
The label 0 is special in that the wavefunctions satisfy
$\tilde{\phi}_0^\ai(x;\bm{\lambda})\tilde{\phi}_0^\ait(x;\bm{\lambda})=
\cF^{-1}\frac{d\eta(x)}{dx}$ since $\xi_0^{\ai, \ait}=1$.
Here the constant $\cF=2$ for L and $\cF=-4$ for J.
We will not use the label 0 states for deletion.

Next we show  that the virtual state solutions at the first step
$\{\tilde{\phi}_{\text{v}}^{(1)}(x)\}$ have no node in the interior.
By using the Schr\"{o}dinger equations for them we obtain
\begin{equation}
  \partial_x\text{W}[\tilde{\phi}_d,\tilde{\phi}_{\text{v}}](x)
  =\bigl(\tilde{\mathcal{E}}_d-\tilde{\mathcal{E}}_{\text{v}}\bigr)
  \tilde{\phi}_d(x)\tilde{\phi}_{\text{v}}(x),
\end{equation}
which has no node. Since  we can verify that
$\text{W}[\tilde{\phi}_d,\tilde{\phi}_{\text{v}}](x)$ vanishes at one
boundary, no-nodeness of $\text{W}[\tilde{\phi}_d,\tilde{\phi}_{\text{v}}](x)$
in the interior follows.
When the virtual states $d$ and $\text{v}$ belong to different types
(I and II) the Wronskians might not vanish at both boundaries.
In that case we can show that they have the same sign at both boundaries
$W[\tilde{\phi}_d,\tilde{\phi}_{\text{v}}](x_1)
W[\tilde{\phi}_d,\tilde{\phi}_{\text{v}}](x_2)>0$.
Likewise we can show, for all the explicit examples in the next section
that $\tilde{\phi}_\text{v}^{(1)}$ and $1/\tilde{\phi}_\text{v}^{(1)}$
have infinite norms.
The steps going from $\mathcal{H}^{(1)}\to \mathcal{H}^{(2)}$ and further
are essentially the same.
A schematic picture illustrating the  virtual state deletion is shown in Fig.3.

\subsubsection{Pseudo virtual state wavefunctions for H, L and J}
\label{pvirtHLJ}

For completeness we list here the pseudo virtual state wave functions 
for the harmonic oscillator (H), the radial oscillator (L) 
and the P\"oschl-Teller (J) potentials:
\begin{align}
\text{H}:& \quad
 \tilde{\phi}_\text{v}\eqdef
  e^{\frac12x^2}\xi_\text{v}(x)),\quad
  \xi_\text{v}(x)\eqdef i^{-\text{v}}H_\text{v}(ix)  \n
  &\quad
  \tilde{\mathcal{E}}_\text{v}\eqdef-2(\text{v}+1)
  =\mathcal{E}\left(-(\text{v}+1)\right),\quad
  \text{v}\in\mathbb{Z}_{\ge0},\quad\tilde{\bm{\delta}}\eqdef-1,
  \label{pvsH}\\
\text{L}:&\quad
  \tilde{\phi}_\text{v}\eqdef
  e^{\frac12x^2}x^{1-g}\xi_\text{v}(\eta(x);g),\quad
  \xi_\text{v}(\eta;g)\eqdef P_\text{v}(-\eta;1-g),\n
  &\quad
  \tilde{\mathcal{E}}_\text{v}\eqdef-4(\text{v}+1)
  =\mathcal{E}\left(-(\text{v}+1)\right),\quad
  \text{v}\in\mathbb{Z}_{\ge0},\quad\tilde{\bm{\delta}}\eqdef-1,
  \label{pvsL}\\
 \text{J}:&\quad
  \tilde{\phi}_\text{v}(x)\eqdef(\sin x)^{1-g}(\cos x)^{1-h}
  \xi_\text{v}(\eta(x);g,h),\quad
  \xi_\text{v}(\eta;g,h)\eqdef P_\text{v}(\eta;1-g,1-h),\n
  &\quad
  \tilde{\mathcal{E}}_\text{v}\eqdef-4(\text{v}+1)
  (g+h-\text{v}-1)=\mathcal{E}\left(-(\text{v}+1)\right),\quad
  \text{v}\in\mathbb{Z}_{\ge0},\quad
  \tilde{\bm{\delta}}\eqdef(-1,-1).
  \label{pvsJ}
\end{align}
\subsection{Multi-indexed Orthogonal Polynomials}
\label{sec:mult-ort}

Here we recapitulate the multi-indexed Laguerre 
and Jacobi polynomials as presented in \cite{os25}.
The basic formula is \eqref{psiM} in  Theorem in \S\ref{sec:darb}.

\begin{center}
  \includegraphics{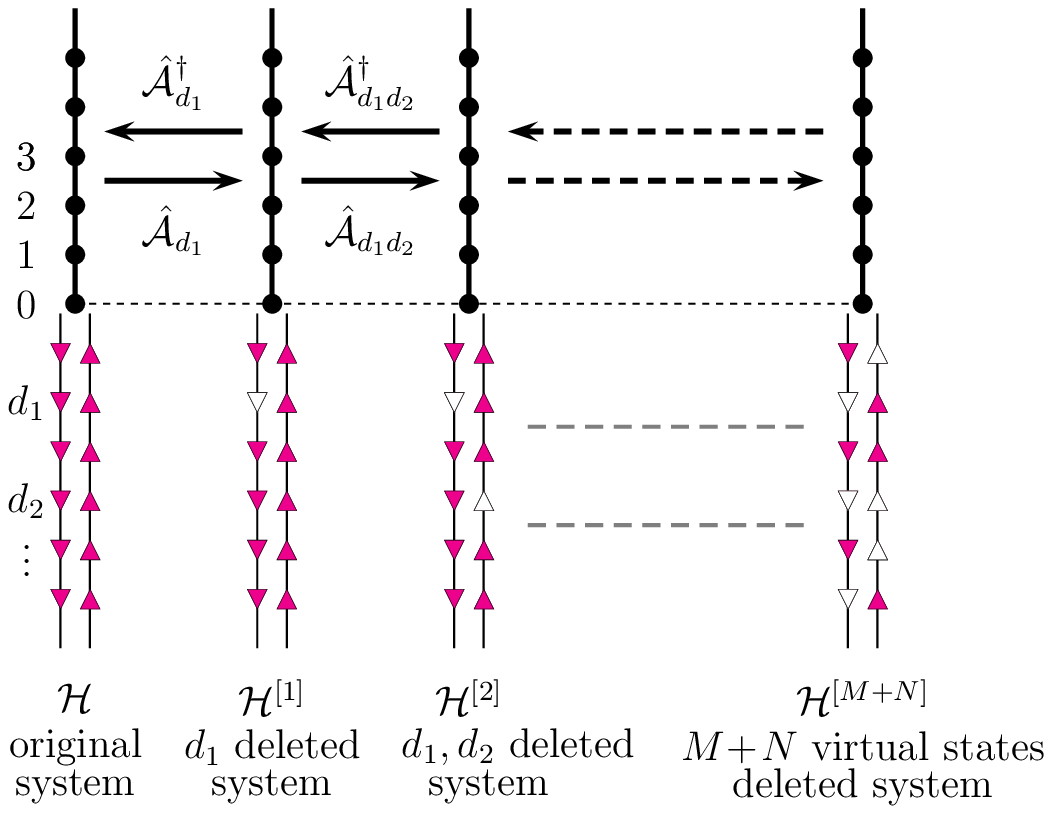}
\end{center}
{\baselineskip=14pt
\begin{quote}
{Figure\,3:  Schematic picture of virtual states deletion.
The black circles denote eigenstates.
The down and up triangles denote virtual states of type I 
and II.
The deleted virtual states are denoted by white triangles.}
\end{quote}
}
Since there are two types of virtual states available, the general deletion
is specified by the set of $M+N$ positive integers
$\mathcal{D}\eqdef\{d_1^\ai,\ldots,d_M^\ai,d_1^\ait,\ldots,d_N^\ait\}$,
$d_j^{\ai,\ait}\ge1$. 
In order to accommodate all these virtual states, the parameters $g$ and
$h$ must be larger than certain bounds:
\begin{align}
  \text{L}:&\quad
  g>\text{max}\{N+\tfrac32,d_j^\ait+\tfrac12\},
  \label{Lbound}\\
  \text{J}:&\quad
  g>\text{max}\{N+2,d_j^\ait+\tfrac12\},\quad
  h>\text{max}\{M+2,d_j^\ai+\tfrac12\}.
  \label{Jbound}
\end{align}

Like the original eigenfunctions \eqref{facteig} the $n$-th eigenfunction
$\phi_{\mathcal{D},n}(x;\bm{\lambda})\equiv\phi_n^{[M,N]}(x)$ after the
deletion \eqref{psiM} can be clearly factorised into an $x$-dependent
part, the common denominator polynomial $\Xi_{\mathcal{D}}$ in $\eta$
and the {\em multi-indexed polynomial} $P_{\mathcal{D},n}$ in $\eta$:
\begin{equation}
  \phi_n^{[M,N]}(x)\equiv\phi_{\mathcal{D},n}(x;\bm{\lambda})
  =\cF^{M+N}\psi_{\mathcal{D}}(x;\bm{\lambda})
  P_{\mathcal{D},n}(\eta(x);\bm{\lambda}),\quad
  \psi_{\mathcal{D}}(x;\bm{\lambda})\eqdef
  \frac{\phi_0(x;\bm{\lambda}^{[M,N]})}
  {\Xi_{\mathcal{D}}(\eta(x);\bm{\lambda})},
\end{equation}
in which $\phi_0(x;\bm{\lambda})$ is the ground state wavefunction
\eqref{ex2}-\eqref{ex3}.
Here the shifted parameters $\bm{\lambda}^{[M,N]}$ after the $[M,N]$
deletion is
$\bm{\lambda}^{[M,N]}\eqdef\bm{\lambda}-M\tilde{\bm{\delta}}^\ai
-N\tilde{\bm{\delta}}^\ait$, explicitly it is
\begin{equation}
  \bm{\lambda}^{[M,N]}=g+M-N\ \ \text{for L},\quad
  \bm{\lambda}^{[M,N]}=(g+M-N,h-M+N)\ \ \text{for J}.
\end{equation}
The polynomials $P_{\mathcal{D},n}$ and $\Xi_{\mathcal{D}}$ are expressed
in terms of Wronskians in the variable $\eta$:
\begin{align}
  P_{\mathcal{D},n}(\eta;\bm{\lambda})&\eqdef
  \text{W}[\mu_1,\ldots,\mu_M,\nu_1,\ldots,\nu_N,P_n](\eta)\n
  &\qquad\times\left\{
  \begin{array}{ll}
  e^{-M\eta}\,\eta^{(M+g+\frac12)N}&:\text{L}\\[2pt]
  \bigl(\frac{1-\eta}{2}\bigr)^{(M+g+\frac12)N}
  \bigl(\frac{1+\eta}{2}\bigr)^{(N+h+\frac12)M}&:\text{J}
  \end{array}\right.,
  \label{multiP}\\
  \Xi_{\mathcal{D}}(\eta;\bm{\lambda})&\eqdef
  \text{W}[\mu_1,\ldots,\mu_M,\nu_1,\ldots,\nu_N](\eta)\n
  &\qquad\times\left\{
  \begin{array}{ll}
  e^{-M\eta}\,\eta^{(M+g-\frac12)N}&:\text{L}\\[2pt]
  \bigl(\frac{1-\eta}{2}\bigr)^{(M+g-\frac12)N}
  \bigl(\frac{1+\eta}{2}\bigr)^{(N+h-\frac12)M}&:\text{J}
  \end{array}\right.,\\
  &\hspace{-15mm}
  \mu_j=\left\{
  \begin{array}{ll}
  e^{\eta}\xi_{d_j^\ai}^\ai(\eta;g)&:\text{L}\\[4pt]
  \bigl(\frac{1+\eta}{2}\bigr)^{\frac12-h}\xi_{d_j^\ai}^\ai(\eta;g,h)
  &:\text{J}
  \end{array}\right.,\quad
  \nu_j=\left\{
  \begin{array}{ll}
  \eta^{\frac12-g}\xi_{d_j^\ait}^\ait(\eta;g)&:\text{L}\\[4pt]
  \bigl(\frac{1-\eta}{2}\bigr)^{\frac12-g}\xi_{d_j^\ait}^\ait(\eta;g,h)
  &:\text{J}
  \end{array}\right.,
\end{align}
in which $P_n$ in \eqref{multiP} denotes the original polynomial,
$P_n(\eta;g)$ for L and $P_n(\eta;g,h)$ for J.
The multi-indexed polynomial $P_{\mathcal{D},n}$ is of degree $\ell+n$
and the denominator polynomial $\Xi_{\mathcal D}$ is of degree $\ell$ in
$\eta$, in which $\ell$ is given by
\begin{equation}
  \ell\eqdef\sum_{j=1}^Md_j^\ai+\sum_{j=1}^Nd_j^\ait
  -\frac12 M(M-1)-\frac12N(N-1)+MN\ge1.
\end{equation}
Here the label $n$ specifies the energy eigenvalue 
$\mathcal{E}(n;\bm{\lambda})$ of
$\phi_{\mathcal{D},n}$ and  it also counts the nodes
 due to the  oscillation theorem in \S\ref{probset}. 
The multi-indexed polynomials $\{P_{\mathcal{D},n}\}$ form a complete set of
orthogonal polynomials with the orthogonality relations:
\begin{align}
  &\quad\int\!\!d\eta\,
  \frac{W(\eta;\bm{\lambda}^{[M,N]})}
  {\Xi_{\mathcal{D}}(\eta;\bm{\lambda})^2}
  P_{\mathcal{D},n}(\eta;\bm{\lambda})
  P_{\mathcal{D},m}(\eta;\bm{\lambda})\n
  &=h_n(\bm{\lambda})\delta_{nm}\times\left\{
  \begin{array}{ll}
  \prod_{j=1}^M(n+g+d_j^\ai+\tfrac12)\cdot
  \prod_{j=1}^N(n+g-d_j^\ait-\tfrac12)&:\text{L}\\[3pt]
  4^{-M-N}
  \prod_{j=1}^M(n+g+d_j^\ai+\tfrac12)(n+h-d_j^\ai-\tfrac12)\\[2pt]
  \hspace{9.4mm}\times
  \prod_{j=1}^N(n+g-d_j^\ait-\tfrac12)(n+h+d_j^\ait+\tfrac12)
  &:\text{J}
  \end{array}\right.,
  \label{multiortho}
\end{align}
where the weight function of the original polynomials
$W(\eta;\bm{\lambda})d\eta=\phi_0(x;\bm{\lambda})^2dx$ reads explicitly
\begin{equation}
  W(\eta;\bm{\lambda})\eqdef
  \left\{\begin{array}{ll}
  \frac12e^{-\eta}\eta^{g-\frac12}&:\text{L}\\
  \frac{1}{2^{g+h+1}}(1-\eta)^{g-\frac12}(1+\eta)^{h-\frac12}&:\text{J}
  \end{array}\right..
\end{equation}

\subsubsection{Deformed Hamiltonians}
\label{sec:defhams}

We explore various properties of the new
multi-indexed polynomials $\{P_{\mathcal{D},n}\}$. 
The lowest
degree polynomial $P_{\mathcal{D},0}(\eta;\bm{\lambda})$ is related to
the denominator polynomial $\Xi_{\mathcal D}(\eta;\bm{\lambda})$ by the
parameter shift $\bm{\lambda}\to\bm{\lambda}+\bm{\delta}$ ($\bm{\delta}=1$
for L and $\bm{\delta}=(1,1)$ for J):
\begin{equation}
  P_{\mathcal{D},0}(\eta;\bm{\lambda})
  =\Xi_{\mathcal{D}}(\eta;\bm{\lambda}+\bm{\delta})\times\left\{
  \begin{array}{ll}
  (-1)^M\prod_{j=1}^N(g-d_j^\ait-\tfrac12)&:\text{L}\\[4pt]
  2^{-M}\prod_{j=1}^M(h-d_j^\ai-\tfrac12)\cdot
  (-2)^{-N}\prod_{j=1}^{N}(g-d_j^\ait-\tfrac12)&:\text{J}
  \end{array}\right..
  \label{plusdelta}
\end{equation}
The Hamiltonian
$\mathcal{H}_{\mathcal D}(\bm{\lambda})\equiv\mathcal{H}^{[M,N]}$ of
the $[M,N]$ deleted system can be expressed in terms of its ground state
eigenfunction $\phi_{\mathcal{D},0}(x;\bm{\lambda})\equiv
\phi_0^{[M,N]}(x;\bm{\lambda})$ with the help of
\eqref{plusdelta}:
\begin{align}
  &\mathcal{H}_{\mathcal D}(\bm{\lambda})
  =\mathcal{A}_{\mathcal D}(\bm{\lambda})^\dagger
  \mathcal{A}_{\mathcal D}(\bm{\lambda}),
  \label{hamD}\\
  &\mathcal{A}_{\mathcal D}(\bm{\lambda})\eqdef
  \frac{d}{dx}-\frac{\partial_x\phi_{\mathcal{D},0}(x;\bm{\lambda})}
  {\phi_{\mathcal{D},0}(x;\bm{\lambda})},\quad
  \phi_{\mathcal{D},0}(x;\bm{\lambda})\propto
  \phi_{0}(x;\bm{\lambda}^{[M,N]})
  \frac{\Xi_{\mathcal D}(\eta(x);\bm{\lambda}+\bm{\delta})}
  {\Xi_{\mathcal D}(\eta(x);\bm{\lambda})}.
  \label{phiD0}
\end{align}
Reflecting the construction \cite{stz,os21} it is shape invariant
\cite{genden,os16,os18}
\begin{equation}
  \mathcal{A}_{\mathcal{D}}(\bm{\lambda})
  \mathcal{A}_{\mathcal{D}}(\bm{\lambda})^{\dagger}
  =\mathcal{A}_{\mathcal{D}}(\bm{\lambda}+\bm{\delta})^{\dagger}
  \mathcal{A}_{\mathcal{D}}(\bm{\lambda}+\bm{\delta})
  +\mathcal{E}(1;\bm{\lambda}).
  \label{shapeinvD}
\end{equation}
This means that the operators 
$\mathcal{A}_{\mathcal{D}}(\bm{\lambda})$ and
$\mathcal{A}_{\mathcal{D}}(\bm{\lambda})^\dagger$ 
relate the eigenfunctions
of neighbouring degrees and parameters:
\begin{align}
  \mathcal{A}_{\mathcal{D}}(\bm{\lambda})
  \phi_{\mathcal{D},n}(x;\bm{\lambda})
  &=f_n(\bm{\lambda})
  \phi_{\mathcal{D},n-1}(x;\bm{\lambda}+\bm{\delta}),
  \label{ADphiDn=}\\
  \mathcal{A}_{\mathcal{D}}(\bm{\lambda})^{\dagger}
  \phi_{\mathcal{D},n-1}(x;\bm{\lambda}+\bm{\delta})
  &=b_{n-1}(\bm{\lambda})
  \phi_{\mathcal{D},n}(x;\bm{\lambda}),
  \label{ADdphiDn=}
\end{align}
in which the constants  
$f_n(\bm{\lambda})$ and $b_{n-1}(\bm{\lambda})$
are the factors of the eigenvalue
$f_n(\bm{\lambda})b_{n-1}(\bm{\lambda})=\mathcal{E}(n;\bm{\lambda})$ 
given in \eqref{fnbn}.
The forward and backward shift operators are defined by
\begin{align}
  \mathcal{F}_{\mathcal{D}}(\bm{\lambda})&\eqdef
  \psi_{\mathcal{D}}\,(x;\bm{\lambda}+\bm{\delta})^{-1}\circ
  \mathcal{A}_{\mathcal{D}}(\bm{\lambda})\circ
  \psi_{\mathcal{D}}\,(x;\bm{\lambda})
  \label{FDdef}\\
  &=\cF\frac{\Xi_{\mathcal{D}}(\eta;\bm{\lambda}+\bm{\delta})}
  {\Xi_{\mathcal{D}}(\eta;\bm{\lambda})}\Bigl(\frac{d}{d\eta}
  -\frac{\partial_{\eta}\Xi_{\mathcal{D}}(\eta;\bm{\lambda}+\bm{\delta})}
  {\Xi_{\mathcal{D}}(\eta;\bm{\lambda}+\bm{\delta})}\Bigr),
  \label{FD}\\
  \mathcal{B}_{\mathcal{D}}(\bm{\lambda})&\eqdef
  \psi_{\mathcal{D}}\,(x;\bm{\lambda})^{-1}\circ
  \mathcal{A}_{\mathcal{D}}(\bm{\lambda})^{\dagger}\circ
  \psi_{\mathcal{D}}\,(x;\bm{\lambda}+\bm{\delta})
  \label{BDdef}\\
  &=-4\cF^{-1}c_2(\eta)\frac{\Xi_{\mathcal{D}}(\eta;\bm{\lambda})}
  {\Xi_{\mathcal{D}}(\eta;\bm{\lambda}+\bm{\delta})}
  \Bigl(\frac{d}{d\eta}
  +\frac{c_1(\eta,\bm{\lambda}^{[M,N]})}{c_2(\eta)}
  -\frac{\partial_{\eta}\Xi_{\mathcal{D}}(\eta;\bm{\lambda})}
  {\Xi_{\mathcal{D}}(\eta;\bm{\lambda})}\Bigr),
  \label{BD}
\end{align}
in which $c_1(\eta;\bm{\lambda})$ and $c_2(\eta)$ 
are given in \eqref{cF,c1,c2}.
Their action on the multi-indexed polynomials
$P_{\mathcal{D},n}(\eta;\bm{\lambda})$ is
\begin{align}
  \mathcal{F}_{\mathcal{D}}(\bm{\lambda})
  P_{\mathcal{D},n}(\eta;\bm{\lambda})
  &=f_n(\bm{\lambda})
  P_{\mathcal{D},n-1}(\eta;\bm{\lambda}+\bm{\delta}),
  \label{FDPDn=}\\
  \mathcal{B}_{\mathcal{D}}(\bm{\lambda})
  P_{\mathcal{D},n-1}(\eta;\bm{\lambda}+\bm{\delta})
  &=b_{n-1}(\bm{\lambda})
  P_{\mathcal{D},n}(\eta;\bm{\lambda}).
  \label{BDPDn=}
\end{align}

\subsubsection{Second Order Equations for the New Polynomials}
\label{sec:secondeq}

The second order differential operator
$\widetilde{\mathcal{H}}_{\mathcal{D}}(\bm{\lambda})$ governing the
multi-indexed polynomials is:
\begin{align}
  \widetilde{\mathcal{H}}_{\mathcal{D}}(\bm{\lambda})
  &\eqdef\psi_{\mathcal{D}}(x;\bm{\lambda})^{-1}\circ
  \mathcal{H}_{\mathcal{D}}(\bm{\lambda})\circ
  \psi_{\mathcal{D}}(x;\bm{\lambda})
  =\mathcal{B}_{\mathcal{D}}(\bm{\lambda})
  \mathcal{F}_{\mathcal{D}}(\bm{\lambda})\n
  &=-4\biggl(c_2(\eta)\frac{d^2}{d\eta^2}
  +\Bigl(c_1(\eta,\bm{\lambda}^{[M,N]})-2c_2(\eta)
  \frac{\partial_{\eta}\Xi_{\mathcal{D}}(\eta;\bm{\lambda})}
  {\Xi_{\mathcal{D}}(\eta;\bm{\lambda})}\Bigr)\frac{d}{d\eta}\n
  &\qquad\quad
  +c_2(\eta)
  \frac{\partial^2_{\eta}\Xi_{\mathcal{D}}(\eta;\bm{\lambda})}
  {\Xi_{\mathcal{D}}(\eta;\bm{\lambda})}
  -c_1(\eta,\bm{\lambda}^{[M,N]}-\bm{\delta})
  \frac{\partial_{\eta}\Xi_{\mathcal{D}}(\eta;\bm{\lambda})}
  {\Xi_{\mathcal{D}}(\eta;\bm{\lambda})}
  \biggr),
  \label{ThamD}\\
  &\hspace{-12mm}\widetilde{\mathcal{H}}_{\mathcal{D}}(\bm{\lambda})
  P_{\mathcal{D},n}(\eta;\bm{\lambda})=\mathcal{E}(n;\bm{\lambda})
  P_{\mathcal{D},n}(\eta;\bm{\lambda}).
  \label{fuchs}
\end{align}
Since all the zeros of $\Xi_{\mathcal{D}}(\eta;\bm{\lambda})$ 
are simple for generic couplings,
\eqref{fuchs} is a Fuchsian differential equation for the J case.
The characteristic exponents at the zeros of
$\Xi_{\mathcal{D}}(\eta;\bm{\lambda})$ are the same everywhere, 0 and 3.
The multi-indexed polynomials $\{P_{\mathcal{D},n}(\eta;\bm{\lambda})\}$
provide infinitely many global solutions of the above Fuchsian equation
\eqref{fuchs} with $3+\ell$ regular singularities for the J case
\cite{hos}. The L case is obtained as a confluent limit. These situations
are basically the same as 
those of the exceptional polynomials.

For special choices of $\mathcal{D}$ and by fine tuning the couplings,  
it is possible to construct a denominator polynomial 
$\Xi_{\mathcal{D}}(\eta;\bm{\lambda})$ 
having higher zeros. Explicit examples of double zeros were constructed 
in \cite{gomez4,st3,hst}. These can be considered as the 
{\em confluence of apparent singularities\/}.

Although we have restricted $d_j^{\ai,\ait}\geq 1$, there is no
obstruction for deletion of $d_j^{\ai,\ait}=0$.
In terms of the multi-indexed polynomial \eqref{multiP}, the level 0
deletions imply the following:
\begin{align}
  &P_{\mathcal{D},n}(\eta;\bm{\lambda})\Bigm|_{d_M^{\ai}=0}
  =P_{\mathcal{D}',n}(\eta;\bm{\lambda}-\tilde{\bm{\delta}}^{\ai})
  \times A,\n
  &\qquad
  \mathcal{D}'=\{d_1^{\ai}-1,\ldots,d_{M-1}^{\ai}-1,
  d_1^{\ait}+1,\ldots,d_N^{\ait}+1\},
  \label{dIM=0}\\
  &P_{\mathcal{D},n}(\eta;\bm{\lambda})\Bigm|_{d_N^{\ait}=0}
  =P_{\mathcal{D}',n}(\eta;\bm{\lambda}-\tilde{\bm{\delta}}^{\ait})
  \times B,\n
  &\qquad
  \mathcal{D}'=\{d_1^{\ai}+1,\ldots,d_M^{\ai}+1,
  d_1^{\ait}-1,\ldots,d_{N-1}^{\ait}-1\},
  \label{dIIN=0}
\end{align}
where the 
multiplicative factors $A$ and $B$ are
\begin{align}
  A&=\left\{
  \begin{array}{ll}
  (-1)^M\prod_{j=1}^N(d_j^{\ait}+1)&:\text{L}\\[4pt]
  -(-2)^{-M}\prod_{j=1}^{M-1}(g-h+d_j^{\ai}+1)\cdot
  (-2)^{-N}\prod_{j=1}^N(d_j^{\ait}+1)\cdot
  (n+h-\frac12)&:\text{J}
  \end{array}\right.,\\
  B&=\left\{
  \begin{array}{ll}
  (-1)^M\prod_{j=1}^M(d_j^{\ai}+1)\cdot
  (n+g-\frac12)&:\text{L}\\[4pt]
  2^{-M}\prod_{j=1}^M(d_j^{\ai}+1)\cdot
  (-2)^{-N}\prod_{j=1}^{N-1}(h-g+d_j^{\ait}+1)\cdot
  (n+g-\frac12)&:\text{J}
  \end{array}\right..
\end{align}
{}From \eqref{plusdelta}, $\Xi_{\mathcal{D}}$ behaves similarly.
Therefore the level 0 deletion corresponds to $M+N-1$ virtual states
deletions. This is why we have restricted $d_j^{\ai,\ait}\ge1$.

These relations \eqref{dIM=0}--\eqref{dIIN=0} can be used for studying
the equivalence of $\mathcal{H}_{\mathcal{D}}$.
See \cite{equi, maya} 
for recent interesting developments on various equivalences.

The exceptional $X_\ell$ orthogonal polynomials of type I and II,
\cite{gomez,quesne,os16,os19,hos,stz} correspond to the simplest cases
of one virtual state deletion of that type, $\mathcal{D}=\{\ell^\ai\}$
or $\{\ell^\ait\}$, $\ell\ge1$:
\begin{equation}
  \xi_{\ell}(\eta;\bm{\lambda})
  =\Xi_{\mathcal{D}}(\eta;\bm{\lambda}+\ell\bm{\delta}+\tilde{\bm{\delta}}),
  \quad
  P_{\ell,n}(\eta;\bm{\lambda})
  =P_{\mathcal{D},n}(\eta;\bm{\lambda}+\ell\bm{\delta}+\tilde{\bm{\delta}})
  \times A,
  \label{excepts}
\end{equation}
where $\tilde{\bm{\delta}}=\tilde{\bm{\delta}}^{\ai,\ait}$ and
the  
multiplicative factor $A$ is $A=-1$ for $X^{\I}L$, $(n+g+\frac12)^{-1}$ 
for $X^{\II}L$,
$2(n+h+\frac12)^{-1}$ for $X^{\I}J$ and $-2(n+g+\frac12)^{-1}$ for 
$X^{\II}J$.
Most formulas between \eqref{hamD} and \eqref{fuchs} look almost the same
as those appearing in the theory of the exceptional orthogonal polynomials
\cite{os16,os19,hos,stz,os21}.

\subsection{Duality between pseudo virtual states and eigenstates}
\label{sec:dual}

For known shape-invariant
potentials,  Darboux transformations in terms of multiple
{\em pseudo virtual state wavefunctions} are  equivalent to
Krein-Adler transformations deleting multiple eigenstates with
{\em shifted parameters}.
See Fig.4 for the illustration.
Let us introduce appropriate symbols 
and notation for stating the duality or equivalence.
As before, let 
$\mathcal{D}\eqdef\{d_1,d_2,\ldots,d_M\}$ 
($d_j\in\mathbb{Z}_{\ge0}$)
be a set of distinct non-negative integers.
We introduce an integer $N$ and fix it to be not less than the maximum of
$\mathcal{D}$:
\begin{equation}
  N\ge\text{max}(\mathcal{D}).
\end{equation}
Let us define another set of distinct non-negative integers
$\bar{\mathcal{D}}=\{0,1,\ldots,N\}\backslash
\{\bar{d}_1,\bar{d}_2,\ldots,\bar{d}_M\}$
together with the shifted parameters $\bar{\bm{\lambda}}$:
\begin{align}
  &\bar{\mathcal{D}}\eqdef\{0,1,\ldots,\breve{\bar{d}}_1,\ldots,
  \breve{\bar{d}}_2,\ldots,\breve{\bar{d}}_M,\ldots,N\}
  =\{e_1,e_2,\ldots,e_{N+1-M}\},\n
  &\bar{d}_j\eqdef N-d_j,\quad
  \bar{\bm{\lambda}}\eqdef \bm{\lambda}-(N+1)\bm{\delta}.
  \label{barD}
\end{align}

\begin{center}
  \scalebox{0.7}{\includegraphics{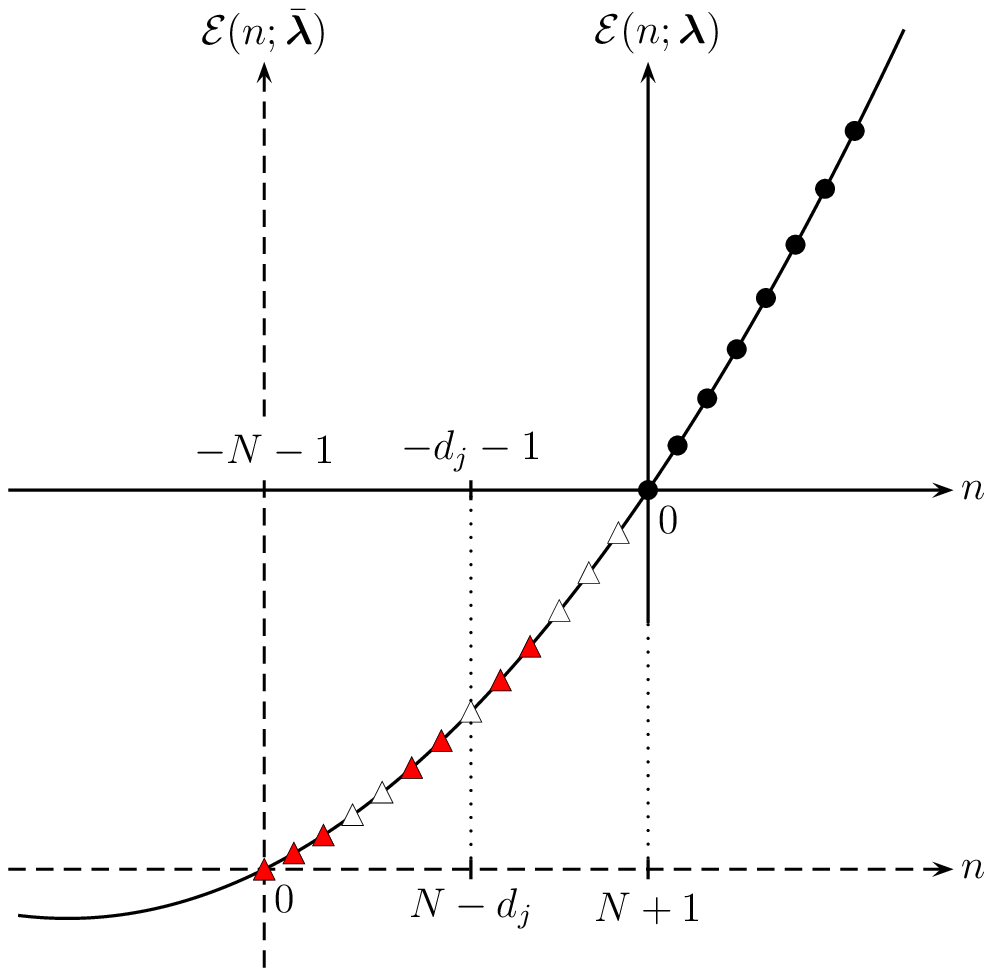}}\quad
  \scalebox{0.7}{\includegraphics{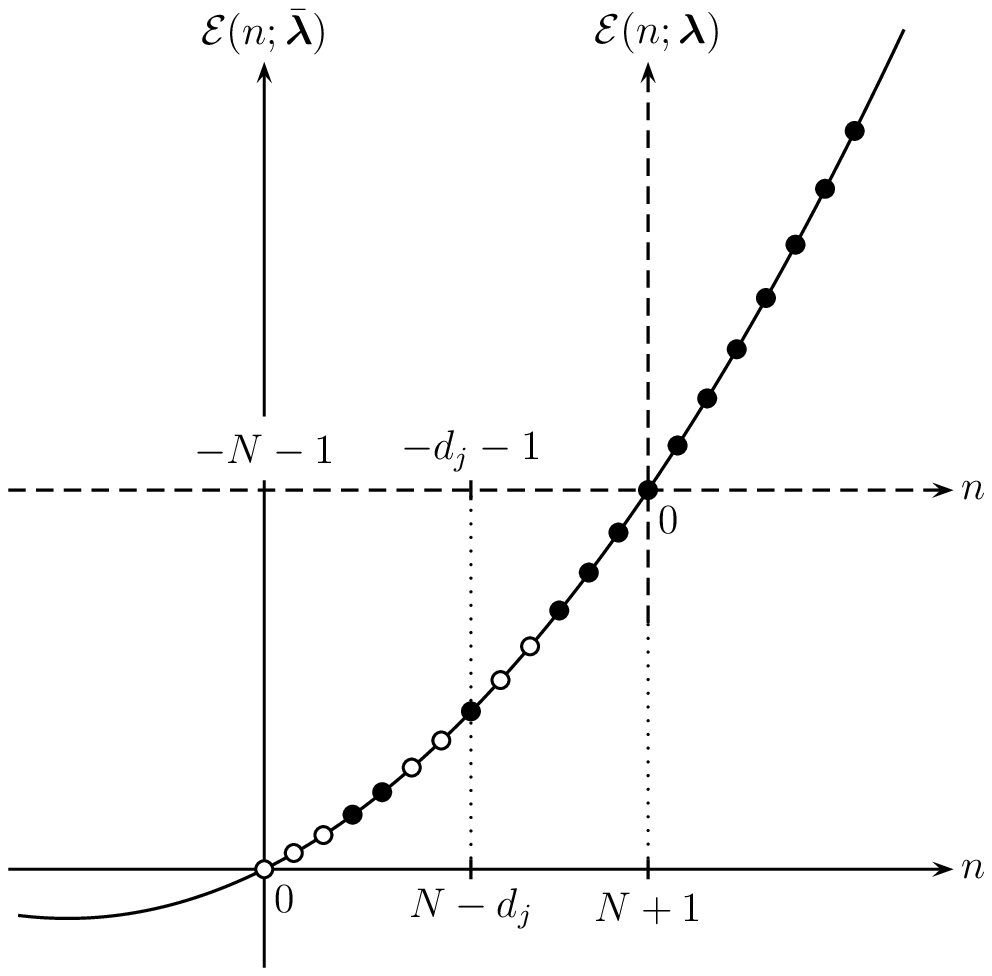}}
 \end{center}
 {\baselineskip=14pt
\begin{quote}
  Figure\ 4:  The left represents the Darboux transformations in terms of
pseudo virtual states. The right corresponds to the Krein-Adler
transformations in terms of eigenstates with shifted parameters.%
The black circles denote eigenstates. 
The white circles in the right graphic denote deleted eigenstates.
The white triangles in the left graphic denote the pseudo virtual states 
used in the Darboux transformations. 
The red triangles denote the unused pseudo virtual states.
\end{quote}
}	

Starting from a shape-invariant original system \eqref{1dham} 
with the parameters $\bm{\lambda}$,
the system after Darboux transformations in terms of a set of pseudo
virtual state wavefunctions $\mathcal{D}$ is described by the Hamiltonian
$\mathcal{H}^{\text{DP}}$
\begin{align}
  &\mathcal{H}^{\text{DP}}=-\frac{d^2}{dx^2}+U^{\text{DP}}(x),\n
  &U^{\text{DP}}(x)=U(x;\bm{\lambda})-2\partial_x^2\log\bigl|
  \text{W}[\tilde{\phi}_{d_1},\tilde{\phi}_{d_2},\ldots,
  \tilde{\phi}_{d_M}](x;\bm{\lambda})\bigr|.
  \label{HDP}
\end{align}
General theory presented in \S\,\ref{sec:darb} states that,
if the Hamiltonian $\mathcal{H}^{\text{DP}}$ is non-singular,
the eigenstates are given by $\Phi^{\text{DP}}_n$ and
$\breve{\Phi}^{\text{DP}}_j$:
\begin{align}
  &\Phi^{\text{DP}}_n(x)=
  \frac{\text{W}[\tilde{\phi}_{d_1},\tilde{\phi}_{d_2},\ldots,
  \tilde{\phi}_{d_M},\phi_n](x;\bm{\lambda})}
  {\text{W}[\tilde{\phi}_{d_1},\tilde{\phi}_{d_2},\ldots,
  \tilde{\phi}_{d_M}](x;\bm{\lambda})}
  \quad(n=0,1,\ldots,),\n
  &\breve{\Phi}^{\text{DP}}_j(x)=
  \frac{\text{W}[\tilde{\phi}_{d_1},\tilde{\phi}_{d_2},\ldots,
  \breve{\tilde{\phi}}_{d_j},\ldots,\tilde{\phi}_{d_M}](x;\bm{\lambda})}
  {\text{W}[\tilde{\phi}_{d_1},\tilde{\phi}_{d_2},\ldots,
  \tilde{\phi}_{d_M}](x;\bm{\lambda})}
  \quad(j=1,2,\ldots,M),\n
  &\mathcal{H}^{\text{DP}}\Phi^{\text{DP}}_n(x)
  =\mathcal{E}(n;\bm{\lambda})\Phi^{\text{DP}}_n(x),\quad
  \mathcal{H}^{\text{DP}}\breve{\Phi}^{\text{DP}}_j(x)
  =\mathcal{E}(-d_j-1;\bm{\lambda})\breve{\Phi}^{\text{DP}}_j(x).
  \label{DPdiffeq}
\end{align}
The system after Krein-Adler transformations 
in terms of $\bar{\mathcal{D}}$
with shifted parameters $\bar{\bm{\lambda}}$ 
is described by the Hamiltonian
$\mathcal{H}^{\text{KA}}$
\begin{align}
  &\mathcal{H}^{\text{KA}}=-\frac{d^2}{dx^2}+U^{\text{KA}}(x),\n
  &U^{\text{KA}}(x)=U(x;\bar{\bm{\lambda}})-2\partial_x^2\log\bigl|
  \text{W}[\phi_0,\phi_1,\ldots,\breve{\phi}_{\bar{d}_1},\ldots,
  \breve{\phi}_{\bar{d}_M},\ldots,\phi_N](x;\bar{\bm{\lambda}})\bigr|.
  \label{HKA}
\end{align}
Here we assume that the original system \eqref{1dham} with the shifted
parameters $\bar{\bm{\lambda}}$ has square integrable eigenstates, etc.
If the Krein-Adler conditions are fulfilled, eigenstates are given by
$\Phi^{\text{KA}}_n$ and $\breve{\Phi}^{\text{KA}}_j$:
\begin{align}
  &\Phi^{\text{KA}}_n(x)=
  \frac{\text{W}[\phi_0,\phi_1,\ldots,\breve{\phi}_{\bar{d}_1},
  \ldots,\breve{\phi}_{\bar{d}_M},\ldots,\phi_N,\phi_{N+1+n}]
  (x;\bar{\bm{\lambda}})}
  {\text{W}[\phi_0,\phi_1,\ldots,\breve{\phi}_{\bar{d}_1},\ldots,
  \breve{\phi}_{\bar{d}_M},\ldots,\phi_N](x;\bar{\bm{\lambda}})}
  \quad(n=0,1,\ldots,),\n
  &\breve{\Phi}^{\text{KA}}_j(x)=
  \frac{\text{W}[\phi_0,\phi_1,\ldots,\breve{\phi}_{\bar{d}_1},
  \ldots,\phi_{\bar{d}_j},\ldots,\breve{\phi}_{\bar{d}_M},\ldots,\phi_N]
  (x;\bar{\bm{\lambda}})}
  {\text{W}[\phi_0,\phi_1,\ldots,\breve{\phi}_{\bar{d}_1},\ldots,
  \breve{\phi}_{\bar{d}_M},\ldots,\phi_N](x;\bar{\bm{\lambda}})}
  \quad(j=1,2,\ldots,M),\n
  &\mathcal{H}^{\text{KA}}\Phi^{\text{KA}}_n(x)
  =\mathcal{E}(N+1+n;\bar{\bm{\lambda}})\Phi^{\text{KA}}_n(x),\quad
  \mathcal{H}^{\text{KA}}\breve{\Phi}^{\text{KA}}_j(x)
  =\mathcal{E}(\bar{d}_j;\bar{\bm{\lambda}})\breve{\Phi}^{\text{KA}}_j(x).
  \label{KAdiffeq}
\end{align}

The duality or the equivalence is stated as the following
\begin{thm}
The two systems with $\mathcal{H}^{\text{\rm DP}}$ and
$\mathcal{H}^{\text{\rm KA}}$ are equivalent.
To be more specific, the equality 
of the potentials and the eigenfunctions
read\/{\rm{:}}
\begin{align}
  U^{\text{\rm DP}}(x)-\mathcal{E}(-N-1;\bm{\lambda})
  &=U^{\text{\rm KA}}(x),
  \label{poteq}\\
  \Phi^{\text{\rm DP}}_n(x)&\propto\Phi^{\text{\rm KA}}_n(x)
  \quad(n=0,1,\ldots,),
  \label{eigeq1}\\
  \breve{\Phi}^{\text{\rm DP}}_j(x)&
  \propto\breve{\Phi}^{\text{\rm KA}}_j(x)
  \quad(j=1,2,\ldots,M).
  \label{eigeq2}
\end{align}
The singularity free conditions of the potential are
\begin{equation}
  \prod_{j=1}^{N+1-M}(m-e_j)\ge0
  \quad(\,\forall m\in\mathbb{Z}_{\geq 0}).
  \label{non-sing}
\end{equation}
For $M=1$, $\mathcal{D}=\{d_1\}$,
$\bar{\mathcal{D}}=\{0,1,\ldots,\breve{\bar{d}}_1,\ldots,N\}$,
the above conditions are satisfied by even $d_1$,
$d_1\in 2\mathbb{Z}_{\geq 0}$.
In other words, the pseudo virtual state wavefunctions
$\{\tilde{\phi}_{\text{\rm v}}\}$ 
for even $\text{\rm v}$ are nodeless.
The above equalities {\rm(}up to multiplicative factors\/{\rm)}
\eqref{poteq}--\eqref{eigeq2} {\rm(}\eqref{eigeq1} with 
$n\in\mathbb{Z}_{\geq 0}$\/{\rm)}
are algebraic and they hold irrespective 
of the non-singularity conditions
\eqref{non-sing}.
\end{thm}
On top of the three fundamental potentials introduced in \S\ref{sec:Exam}, 
the duality holds also for other shape invariant 
and thus exactly solvable potentials; 
Coulomb potential plus the centrifugal barrier, 
Kepler problem in spherical space, 
Morse potential, soliton potential, Rosen-Morse potential,
Hyperbolic symmetric top $\II$, Kepler problem in hyperbolic space, 
hyperbolic P\"{o}schl-Teller potential
\cite{os29}.

We refer to \cite{gos, os29} for proofs and related discussions. 
When the eigenfunctions have the 
factorised form as shown in \eqref{facteig}, 
which is true for H, L and J, 
the following polynomial Wronskian identities hold
\begin{equation}
  \text{W}[\xi_{d_1},\xi_{d_2},\ldots,\xi_{d_M}](\eta;\bm{\lambda})
  \propto
  \text{W}[P_0,P_{1},\ldots,\breve{P}_{\bar{d}_1},\ldots,
  \breve{P}_{\bar{d}_M},\ldots,P_N](\eta;\bar{\bm{\lambda}}).
  \label{genwronide}
\end{equation}
Verifying these identities for various choices 
of $\mathcal{D}=\{d_1,\ldots,d_M\}$ 
for the three examples 
of pseudo virtual state wave functions 
\eqref{pvsH}--\eqref{pvsJ} in \S\ref{pvirtHLJ}
is left to readers as an exercise.
\section{Exactly Solvable Scattering Problems}
\label{sec:Scat}

\subsection{Scattering Problems}
\label{scatpr}

For non-confining potentials, 
when the physical regions extend to infinity, 
the systems have 
continuous spectrum as well as discrete eigenstates.
We adopt the convention that the potentials vanish at infinity,%
\footnote{The cases where $U(-\infty)\neq U(+\infty)$ 
must be treated separately. 
See for example Rosen-Morse potential 
\cite{infhull,susyqm,os28,os29,hls}.
} %
so that the plane waves $e^{\pm ikx}$, $k\in\mathbf{R}_{+}$, 
are the solutions of the Schr\"odinger equation with the 
positive energy $\mathcal{E}=k^2$ in the asymptotic regions.
There are two types,  the full line $x_1=-\infty$, 
$x_2=+\infty$ case called Group (A) or a half line 
$x_1=0$, $x_2=+\infty$ case called Group (B). 
For the continuous spectrum states, 
we consider the scattering problems. 
Fig.5 shows the general situation of the full line
scattering problem.
On top of determining 
the discrete eigenstates \eqref{Sch_eq} as above, 
we need to determine the transmission amplitude 
$t(k)$ (Group (A)) and the
reflection amplitude $r(k)$ (both Group (A) and (B)) 
through the asymptotic behaviours of the 
wave function $\psi_k(x)$:
\begin{alignat}{2}
\mathcal{H}\psi_k(x)&=k^2\psi_k(x), &\quad &k\in\mathbf{R}_{+}, 
\label{psik}\\
\psi_k(x)&\approx\left\{
\begin{array}{cl}
 e^{ikx} &   \qquad x\to+\infty   \\
A(k) e^{ikx}+B(k)e^{-ikx}&  \qquad  x\to-\infty 
\end{array}
\right. &\quad  &\text{(A)},
\label{full}\\
\psi_k(x)&\approx \quad r(k)e^{ikx}+e^{-ikx} 
\qquad  \qquad  \quad   
x\to+\infty ,
&\quad     &\text{(B)}.
\label{half}
\end{alignat}
For the full line scattering (Group (A)) 
one sends a unit amplitude of the right moving wave ($e^{ikx}$) 
at $x=-\infty$. Then a reflected left moving wave 
$e^{-ikx}$ with the amplitude 
$r(k)\eqdef B(k)/A(k)$ is observed at $x=-\infty$
and a transmitted right moving wave  ($e^{ikx}$)  with the amplitude 
$t(k)\eqdef1/A(k)$ is observed at $x=+\infty$.
For the half line scattering (Group (B)), there is no transmitted wave.

\begin{center}
{\includegraphics{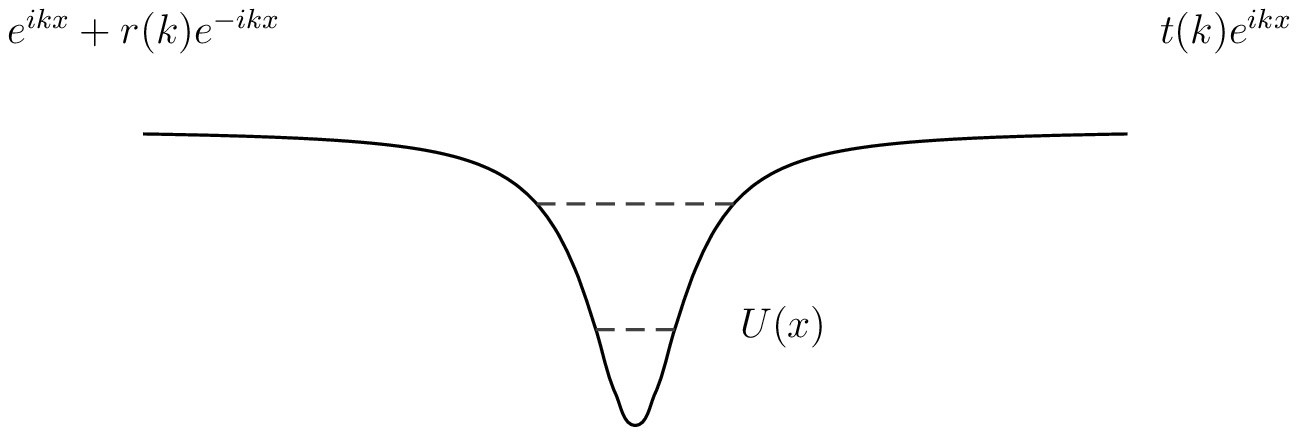}}
 \end{center}
 {\baselineskip=14pt
\begin{quote}
  Figure\ 5:  The general image of a full line scattering 
  problem with the potential $U(x)$. 
  The right going wave  with the unit amplitude 
  $e^{ikx}$ is injected at $-\infty$.
  The reflected  wave $e^{-ikx}$ at $-\infty$ has 
  an amplitude $r(k)$ and the transmitted wave 
  $e^{ikx}$ at $+\infty$ has an amplitude
  $t(k)$. The dashed lines show the discrete eigenlevels.
  \end{quote}
  }
  Since the wave function $\psi_k(x)$ \eqref{psik} is 
  an analytic function of the wave number $k$,
the reflection $r(k)$ and the transmission $t(k)$ amplitudes 
are {\em meromorphic functions\/} of $k$.
The above asymptotic behaviours should be compared 
with those of the discrete eigenstates:
\begin{align}
\mathcal{H}\phi_n(x)&=\mathcal{E}(n)\phi_n(x),\quad 
  \mathcal{E}(0)<\mathcal{E}(1)<\mathcal{E}(2)<\cdots
  <\mathcal{E}(n_{\text{max}})<0,
  \label{finincrease}\\
  \phi_n(x)&\approx c_\pm e^{\mp \sqrt{-\mathcal{E}(n)}\,x},
  \quad x\to\pm\infty.
  \label{eigasym}
\end{align}
By comparing the asymptotic behaviours \eqref{full}, \eqref{half},
 \eqref{eigasym}, 
one finds that the {\em zeros} of $A(k)$, {\em i.e.\/}  
the  {\em poles\/} of the transmission $t(k)$
 and the reflection amplitude $r(k)$ 
 on the {\em positive imaginary axis}  
 $k=i\kappa$, $\kappa\in\mathbf{R}_{+}$
correspond to the discrete spectrum:
\begin{equation}
r(k)\approx \frac{const}{k-i\kappa},\quad 
\kappa\in\mathbf{R}_{+},\quad 
 \exists n\in\{0,\ldots,n_{\text{max}}\},
\quad -\kappa^2=\mathcal{E}(n).
\end{equation}

The scattering amplitudes  of shape invariant systems satisfy 
the constraints of shape invariance \cite{hls,KS}:
\begin{align} 
&\text{full line}:\ t(k;\bm{\lambda}+\bm{\delta})
=\left(\frac{ik+W_+}{ik+W_-}\right) t(k;\bm{\lambda}),\ 
r(k;\bm{\lambda}+\bm{\delta})
=\left(\frac{-ik+W_-}{ik+W_-}\right) r(k;\bm{\lambda}),\\
&\text{half line}:\ r(k;\bm{\lambda}+\bm{\delta})
=\left(\frac{ik+W_+}{-ik+W_+}\right) r(k;\bm{\lambda}),
\label{halfshape}\\
&\qquad \qquad \ \ W_+\eqdef-\lim_{x\to+\infty}
\frac{\partial_x\phi_0(x;\bm{\lambda})}{\phi_0(x;\bm{\lambda})},\qquad
W_-\eqdef-\lim_{x\to-\infty}
\frac{\partial_x\phi_0(x;\bm{\lambda})}{\phi_0(x;\bm{\lambda})}.
\end{align}
These are simply obtained by evaluating the shape invariance relation
\begin{equation*}
\psi_k(x;\bm{\lambda}+\bm{\delta})\propto\psi_k^{(1)}(x;\bm{\lambda})
=\left(\frac{d}{dx}-
\frac{\partial_x\phi_0(x;\bm{\lambda})}{\phi_0(x;\bm{\lambda})}
\right)\psi_k(x;\bm{\lambda})
\end{equation*}
asymptotically.

%
%
\subsection{Reflectionless Potentials}
\label{sec:refl}
Reflectionless potentials of Schr\"odinger equations \cite{KM} 
played a very important role in theoretical physics.
With  special time dependence, 
they describe soliton solutions \cite{hirota} 
of Korteweg de Vries (KdV) equation.
As can be easily understood reflectionless potentials 
must be everywhere negative. 
Let us consider a reflectionless potential $U_N(x)$, 
which has $N$ discrete eigenstates:
\begin{align}
  &\mathcal{H}=-\frac{d^2}{dx^2}+U_N(x),\qquad 
  \mathcal{H}\psi_k(x)=k^2 \psi_k(x),
  \label{schr}\\
  &\mathcal{H}\phi_{N,j}(x)=
  \mathcal{E}_j\phi_{N,j}(x),\quad 
  \mathcal{E}_j=-k_j^2,
  \quad j=1,\ldots,N,\quad 0<k_1<k_2<\cdots<k_N.
    \label{sheqN}
\end{align}
According to Kay and Moses \cite{KM}, it has an expression 
\begin{align}
 U_N(x)&\eqdef-2\partial_x^2\log u_N(x),
 \label{logot}\\
u_N(x)&\eqdef\det A_N(x),\quad
(A_N(x))_{m\,n}\eqdef\delta_{m\,n}
+\frac{c_m e^{-(k_m+k_n)x}}{k_m+k_n},
\quad m,n=1,\ldots,N,
\label{logot2}
\end{align}
in which $\{c_m\}$ are arbitrary positive parameters.
A special choice of $t$-dependence of $\{c_m\}$
\begin{equation*}
c_j\to c_je^{8k_j^3t},\quad j=1,\ldots,N,
\end{equation*}
changes the reflectionless potential $U_N(x)$ to 
an $N$-{\em soliton solution\/} $U_N(x;t)$ 
of the KdV equation \cite{hirota}:
\begin{align*}
 U_N(x;t)&\eqdef-2\partial_x^2\log u_N(x;t), \\
u_N(x;t)&\eqdef\det A_N(x;t),\quad
(A_N(x;t))_{m\,n}\eqdef\delta_{m\,n}
+\frac{c_m e^{-(k_m+k_n)x+8k_m^3t}}{k_m+k_n},\\
0&=\partial_tU_N-6U_N\partial_xU_N+\partial_x^3U_N.
\end{align*}

The very form of  $U_N(x)$ \eqref{logot} suggests 
that {\em the reflectionless potential can be obtained from 
the  trivial potential $U(x)\equiv0$ 
by multiple Darboux transformations\/} \cite{refl}.
The Schr\"odinger equation with $U\equiv0$ has 
{\em square non-integrable solutions\/}
\begin{align}
&\psi_j(x)\eqdef e^{k_jx}+{\tilde{c}_j}e^{-k_jx},
\quad 0<k_1<k_2<\cdots<k_N,\quad (-1)^{j-1}\tilde{c}_j>0,
\label{cjsign}\\
&-\partial_x^2\psi_j(x)=-k_j^2\psi_j(x),\quad j=1,\ldots,N.
\end{align}
The inverses $\{1/\psi_j(x)\}$ 
are locally square integrable at $x=\pm\infty$.
With the above sign of the parameters 
$\{\tilde{c}_j\}$ \eqref{cjsign}  the Wronskian 
of these seed solutions $\{\psi_j\}$ 
$\text{W}[\psi_1,\cdots,\psi_N](x)$ 
is positive and it gives $u_N(x)$ 
up to a factor which is annihilated by $\partial_x^2$
after taking the logarithm:
\begin{align}
&\text{W}[\psi_1,\cdots,\psi_N](x)
=\prod_{j>l}^N(k_j-k_l)\cdot e^{\sum_{j=1}^Nk_jx}u_N(x),\\
&U_N(x)=-2\partial_x^2\log\text{W}[\psi_1,\cdots,\psi_N](x)
=-2\partial_x^2\log u_N(x).
\label{darbN}
\end{align}
Here we have redefined the coefficient of 
$e^{-2k_jx}$ in $u_N(x)$ to be $c_j/(2k_j)$, $c_j>0$.
Similar derivation of the reflectionless potential, 
without the eigenfunctions,  was reported
more than twenty years ago \cite{matv-sal}.
As explained in \S\ref{sec:darb} \eqref{varphiM} 
the above constructed $U_N(x)$ \eqref{darbN} 
has $N$-discrete eigenvalues 
$\mathcal{E}_j=-k_j^2$ with the eigenfunctions
\begin{equation}
\phi_{N,j}(x)\propto \frac{\text{W}[\psi_1,\cdots,
\breve{\psi}_j,\cdots,\psi_N](x)}
{\text{W}[\psi_1,\cdots,\psi_N](x)},\quad 
j=1,\ldots,N.
\end{equation}
By the same multiple Darboux transformation, 
the right moving plane wave solution
$e^{ikx}$ ($k>0$) of the $U\equiv0$
Schr\"odinger equation is mapped to
\begin{align}
e^{ikx}\to \frac{\text{W}[\psi_1,\cdots,\psi_N,e^{ikx}](x)}
{\text{W}[\psi_1,\cdots,\psi_N](x)}
\sim\left\{
\begin{array}{ccc}
\prod_{j=1}^N(ik-k_j)\cdot e^{ikx}  &  x\to +\infty\\[4pt]
\prod_{j=1}^N(ik+k_j)\cdot e^{ikx}    &   x\to -\infty
\end{array}
\right.,
\label{reflplane}
\end{align}
as the Wronskian of  exponential functions is a van der Monde determinant:
\begin{equation}
\text{W}[e^{\alpha_1x},e^{\alpha_2x},\ldots,e^{\alpha_M x}](x)=
\prod_{1\le k<j\le M}(\alpha_j-\alpha_k)\cdot 
e^{\sum_{j=1}^M \alpha_j x}.
\label{vdMonde}
\end{equation}
This scattering wave solution   has reflectionless asymptotic behaviour, 
which is consistent with \eqref{full} with $B(k)\equiv0$.
This is an alternative  derivation of the reflection potential \eqref{logot}.  
{\em Its reflectionless property  and the exact solvability 
are quite  intuitively understood\/}.

%
%
\subsection{Extensions of Solvable Scattering Problems}
\label{sec:exten}

Here we discuss extensions (deformations) of 
solvable scattering problems through
multiple Darboux transformations in terms of polynomial type seed solutions
$\{\tilde{\phi}_{d_j}(x)\}$, $j=1,\ldots,M$, 
indexed by a set of non-negative integers
$\mathcal{D}=\{d_1,\ldots,d_M\}$ which are the 
{\em degrees\/} of the polynomial parts of the seed solutions.
The asymptotic behaviours of the polynomial type seed solution 
$\tilde{\phi}_\text{v}(x)$ are characterised by the
{\em asymptotic exponents\/} $\Delta_\text{v}^\pm$:
\begin{align} 
 \tilde{\phi}_\text{v}(x)&\approx
 \left\{
\begin{array}{cl}
e^{x\Delta_\text{v}^+} &  \qquad  x\to+\infty   \\
e^{x\Delta_\text{v}^-}&   \qquad x\to-\infty 
\end{array}
\right.\quad \text{(A)}, 
\label{asymfulls}\\
\tilde{\phi}_\text{v}(x)&\approx \quad 
\ e^{x\Delta_\text{v}^+}  
\quad  \quad  \ \  x\to+\infty, \quad 
\ \  \text{(B)}.\label{asymcouls}
\end{align}
The  extensions by multiple Darboux transformations 
in terms of polynomial type seed solutions 
$\{\tilde{\phi}_{d_j}(x)\}$, $j=1,\ldots,M$ have been given in 
{\bf Theorem\/} in \S\ref{sec:darb}, 
\eqref{scheq3}--\eqref{varphiM}.
One has to make sure  that the 
{\em deformed potential is non-singular\/}.  
That requires the condition that  the Wronskian 
$\text{W}[\tilde{\phi}_{d_1},\tilde{\phi}_{d_2},\ldots, 
\tilde{\phi}_{d_M}](x)$
{\em should not have any zeros in the interval\/} 
$-\infty<x<\infty$ (A), or $0<x<\infty$ (B).
For the discrete eigenstates $\{\phi_{\mathcal{D},\,n}^{(M)}(x)\}$ 
and the scattering states 
$\{\psi_{\mathcal{D},\,k}^{(M)}(x)\}$ the transformation is iso-spectral.
We stress, however, that additional discrete eigenstates 
may be created below the original ground state level $\mathcal{E}(0)$. 
Their number is equal to that of the used 
{\em pseudo virtual state wavefunctions\/}.

The deformation potential $-2\partial_x^2\log
  \bigl|\text{W}[\tilde{\phi}_{d_1},\tilde{\phi}_{d_2},\ldots,
  \tilde{\phi}_{d_M}](x)\bigr|$ vanishes asymptotically, $x\to\pm\infty$
for the  seed solutions \eqref{asymfulls}--\eqref{asymcouls}.
The deformed continuous spectrum also starts at $\mathcal{E}=0$ 
and the relationship between
the energy $\mathcal{E}$ and the wave number 
$k$, $\mathcal{E}=k^2$ is unchanged.
The multi-indexed scattering amplitudes are easily obtained 
from the asymptotic form of the
wavefunction $\psi_{\mathcal{D},\,k}^{(M)}(x)$ \eqref{varphiM} 
by using the asymptotic forms of the original 
wavefunction $\psi_k(x)$ \eqref{full}--\eqref{half} 
and those of the polynomial seed solutions 
$\tilde{\phi}_\text{v}(x)$ \eqref{asymfulls}--\eqref{asymcouls}.
For the full line scattering (Group (A)) case, we obtain
\begin{align} 
\psi_{\mathcal{D},\,k}^{(M)}(x)&\approx 
\prod_{j=1}^M (ik-\Delta_{d_j}^+)\cdot e^{ikx} 
\hspace{71mm} x\to+\infty, \\
 \psi_{\mathcal{D},\,k}^{(M)}(x)&\approx 
 \prod_{j=1}^M (ik-\Delta_{d_j}^-)\cdot A(k)\,e^{ikx}
 + \prod_{j=1}^M (-ik-\Delta_{d_j}^-)\cdot 
 B(k)\,e^{-ikx}\qquad x\to-\infty,  
 \label{fullleft}
\end{align}
which lead to multi-indexed multiplicative deformations 
of the transmission and reflection amplitudes:
\begin{equation}
\text{(A)}:\quad
t_{\mathcal D}(k)=\prod_{j=1}^M
\frac{k+i\Delta_{d_j}^+}{k+i\Delta_{d_j}^-}\cdot t(k),\quad
r_{\mathcal D}(k)=(-1)^M\prod_{j=1}^M
\frac{k-i\Delta_{d_j}^-}{k+i\Delta_{d_j}^-}\cdot r(k).
\label{trDfull}
\end{equation}
For the half line scattering (Group (B)) case, 
similar calculation gives
\begin{align} 
& \text{(B)}:\quad \psi_{\mathcal{D},\,k}^{(M)}(x)\approx
 \prod_{j=1}^M (ik-\Delta_{d_j}^+)\cdot r(k)\,e^{ikx}
 + \prod_{j=1}^M (-ik-\Delta_{d_j}^+)\cdot\,
 e^{-ikx}\qquad x\to+\infty,  
 \label{defhalf}\\
 &\qquad  \text{(B)}:\quad
r_{\mathcal D}(k)=(-1)^M\prod_{j=1}^M
\frac{k+i\Delta_{d_j}^+}{k-i\Delta_{d_j}^+}\cdot r(k).
\label{rBC}
\end{align}
The meromorphic character of the scattering amplitudes is 
preserved by the multi-indexed extensions.  
The added poles and zeros all appear on the imaginary $k$-axis 
determined solely by the asymptotic exponents of 
the used polynomial type seed solutions.
The derivation depends on the simple fact:
the Wronskian of exponential functions 
$\text{W}[e^{\alpha_1x},e^{\alpha_2x},\ldots,e^{\alpha_M x}](x)$
is  reduced to a van der Monde determinant \eqref{vdMonde},
and most factors cancel out between the numerator 
and denominator of \eqref{psiM}.

%
%
\subsection{Example: Soliton Potential}
\label{sec:scatexa}

Here we  present one typical example of shape invariant 
and solvable potentials.
That is, the soliton potential.
For more examples, see \cite{hls}.
The useful data are
the eigenenergies, eigenfunctions, scattering data, {\em i.e,\/} 
the transmission and reflection amplitudes,
various polynomial seed solutions, 
the virtual and pseudo virtual state wavefunctions, 
the overshoot eigenfunctions together 
with the corresponding asymptotic exponents.

The soliton potential system has finitely many discrete eigenstates
$0\le n\le n_\text{max}(\bm{\lambda})=[h]'$ 
in the specified parameter range:
\begin{align}
  &\bm{\lambda}=h,\quad \bm{\delta}=-1,\quad
  -\infty<x<\infty,\quad h>1/2,
     \label{solpot}\\
&U(x;h)=-\frac{h(h+1)}{\cosh^2x},\quad 
\mathcal{E}(n;h)=-(h-n)^2,\quad \eta(x)=\tanh x,\\
  &\phi_n(x;h)=(\cosh x)^{-h+n}\times 
  P^{(h-n,h-n)}_n(\tanh x),\quad W_+=-W_-=h.
  \label{soleig}
\end{align}
The transmission and reflection amplitudes are:
\begin{align}
t(k;h)&=\frac{\Gamma(-h-ik)\Gamma(1+h-ik)}{\Gamma(-ik)\Gamma(1-ik )},
\quad 
r(k;h) =  
\frac{\Gamma(ik)\Gamma(-h-ik)\Gamma(1+h-ik)}{\Gamma(-ik)\Gamma(-h)
\Gamma(1+h)}.
\label{solt}
\end{align}
The poles on the positive imaginary $k$-axis coming from 
the first Gamma function factor in the numerator 
of $t(k;h)$ \eqref{solt}, $-h-i{k}=-n$, $\Rightarrow k=i(h-n)$,  
$n=0,1,\ldots, [h]'$ provide the eigenspectrum as above. 
As is well known, 
at {\em integer $h=N\in\mathbb{Z}_{\ge1}$ the reflectionless potential 
$r(k;h)\equiv0$ is realised\/} 
by the poles of the Gamma function $\Gamma(-N)$ 
in the denominator of $r(k;h)$. 
In fact, $U_N(x)=-N(N+1)/\cosh^2x$ is a very special case
of the generic reflectionless potential \eqref{logot}--\eqref{logot2} 
for the choice of parameters
\begin{equation}
k_j=j, \quad c_j=\frac{(N+j)!}{j!(j-1)!(N-j)!},\quad j=1,\ldots,N.
\end{equation}
The potential and the scattering amplitudes \eqref{solt}
 are invariant under the discrete transformation $h\to -(h+1)$, 
but the eigenvalues and the eigenfunctions are not.
The relation $|t(k;h)|^2+|r(k;h)|^2=1$, $k\in\mathbf{R}_+$ holds.
In this particular example, 
the scattering data $r(k;h)$ and $t(k;h)$ can be obtained by 
analytically continuing the eigenfunction 
$\phi_n(x;h)$ \eqref{soleig} through
$i k\eqdef-h+n$, by rewriting the Jacobi polynomial 
$P_n^{(h-n,h-n)}(\tanh x)$ in terms of 
the Gauss hypergeometric function \eqref{defL} 
and using its connection formulas.

\paragraph{Polynomial type seed solutions}

The discrete symmetry $h\to-h-1$ generates 
the pseudo virtual  wavefunctions, 
which lie below the ground state:
\begin{align}
  &\text{pseudo virtual}:\ \tilde{\phi}_{\text{v}}(x;h)
  =(\cosh x)^{h+1+\text{v}}
  P_{\text{v}}^{(-h-1-\text{v},-h-1-\text{v})}(\tanh x)
  \ (\text{v}\in\mathbb{Z}_{\ge0}),\n
&\Delta_{\text{v}}^+=h+1+\text{v}>0,\quad 
\Delta_{\text{v}}^-=-\Delta_{\text{v}}^+<0, \quad
  \tilde{\mathcal{E}}_{\text{v}}(h)
    =\mathcal{E}({-\text{v}-1};h)<\mathcal{E}(0;h).
    \label{solpv}
\end{align}
The overshoot eigenfunctions \cite{quesne12} 
provide `pseudo' virtual state  wavefunctions for this potential
 for $\text{v}>2h$:
\begin{align}
  &\text{`pseudo virtual'}:\quad 
  \tilde{\phi}^{\text{os}}_{\text{v}}(x;h)
  =\phi_{\text{v}}(x;h),\n
&\Delta_{\text{v}}^+=-h+\text{v}>0,\quad 
\Delta_{\text{v}}^-=-\Delta_{\text{v}}^+<0,
\quad
  \tilde{\mathcal{E}}^{\text{os}}_{\text{v}}(h)
  <\mathcal{E}(0;h) \quad(\text{v}>2h).
      \label{solpv2}
\end{align}

\paragraph{Deformed scatterings}
A pseudo (`pseudo') virtual state wavefunction will add 
a new discrete eigenstate at its energy.
It is trivial to verify that the pseudo \eqref{solpv} 
and `pseudo' \eqref{solpv2} virtual wavefunctions will add a pole 
on the positive imaginary $k$-axis at $k=i(\text{v}+h+1)$, 
$k=i(\text{v}-h)$, respectively,
 with exactly the same energy of the employed seed solution,
$-(h+\text{v}+1)^2$ and $-(\text{v}-h)^2$, respectively.
For both the pseudo \eqref{solpv} 
and `pseudo' \eqref{solpv2} virtual wavefunctions, 
$\Delta_\text{v}^+=-\Delta_\text{v}^-$. 
This means that  the deformation factors of the transmission 
and reflection amplitudes are  the same except for 
a sign $(-1)^M$:
\begin{equation}
\frac{t_{\mathcal D}(k;h)}{t(k;h)}
=(-1)^M\frac{r_{\mathcal D}(k;h)}{r(k;h)}=
\prod_{j=1}^M\frac{k-i\Delta_{d_j}^-}{k+i\Delta_{d_j}^-}.
\label{defrt}
\end{equation}

\section{Summary and Comments}
\label{sec:Summ}

The basic structure and the recent developments 
in the theory of exactly solvable quantum
mechanics are presented in an elementary way. 
Exactly solvable multi-particle dynamics, in particular, 
Calogero-Moser systems
based on various root systems \cite{Cal-Sut}--\cite{kps} 
could not be included.
The concepts and methods of solvable quantum mechanics,
the factorised Hamiltonians, Crum's theorem and its modifications, 
generic Darboux transformations,
shape invariance, Heisenberg operator solutions, 
rational extensions in terms of 
polynomial type seed solutions, etc, 
can be  generalised to {\em discrete quantum mechanics\/}
\cite{os24}--\cite{os13}, \cite{os14}, 
in which Schr\"odinger equations are {\em difference equations\/}.
We strongly believe that 
these new progress would be interesting to most readers.
\section*{Acknowledgements}

R.\,S. thanks Pauchy Hwang, NTU,  for the invitation  
to  write this review and for timely prompting.

\section*{Appendix: Symbols, Definitions \& Formulas}
\label{append}
\renewcommand{\theequation}{A.\arabic{equation}}

$\circ$ shifted factorial (Pochhammer symbol) $(a)_n$ :
\begin{equation}
   (a)_n\eqdef\prod_{k=1}^n(a+k-1)=a(a+1)\cdots(a+n-1)
   =\frac{\Gamma(a+n)}{\Gamma(a)}.
   \label{defPoch}
\end{equation}
$\circ$ hypergeometric series ${}_rF_s$ :
\begin{equation}
   {}_rF_s\Bigl(\genfrac{}{}{0pt}{}
   {a_1,\,\cdots,a_r}{b_1,\,\cdots,b_s}
   \Bigm|z\Bigr)
   \eqdef\sum_{n=0}^{\infty}
   \frac{(a_1,\,\cdots,a_r)_n}{(b_1,\,\cdots,b_s)_n}\frac{z^n}{n!}\,,
   \label{defhypergeom}
\end{equation}
where $(a_1,\,\cdots,a_r)_n\eqdef\prod_{j=1}^r(a_j)_n
=(a_1)_n\cdots(a_r)_n$.\\
$\circ$ differential equations
\begin{align}
\text{H}:\quad& 
\partial^2_xH_n(x)-2x \partial_xH_n(x)+2nH_n(x)=0,
\label{Hdifeq}\\
\text{L}:\quad& x\partial_x^2L_n^{(\alpha)}(x)
 +(\alpha+1-x)\partial_xL_n^{(\alpha)}(x)
 +nL_n^{(\alpha)}(x)=0,
 \label{Ldiffeq}\\
\text{J}:\quad&
(1-x^2)\partial_x^2P_n^{(\alpha,\beta)}(x)
 +\bigl(\beta-\alpha-(\alpha+\beta+2)x\bigr)
 \partial_x P_n^{(\alpha,\beta)}(x)\n
 &\qquad \qquad \qquad 
 +n(n+\alpha+\beta+1)P_n^{(\alpha,\beta)}(x)=0.
 \label{Jdiffeq}
\end{align}
$\circ$ Rodrigues formulas
\begin{align}
\text{H}:\quad& H_n(x)=(-1)^n 
e^{x^2}\Bigl(\frac{d}{dx}\Bigr)^ne^{-x^2},
\label{HRod}\\
\text{L}:\quad& L_n^{(\alpha)}(x)=
\frac{1}{n!}\frac{1}{e^{-x}x^{\alpha}}
 \Bigl(\frac{d}{dx}\Bigr)^n\bigl(e^{-x}x^{n+\alpha}\bigr).
 \label{LRod}\\
\text{J}:\quad&
 P_n^{(\alpha,\beta)}(x)=\frac{(-1)^n}{2^nn!}
 \frac{1}{(1-x)^{\alpha}(1+x)^{\beta}}
 \Bigl(\frac{d}{dx}\Bigr)^n\bigl((1-x)^{n+\alpha}(1+x)^{n+\beta}\bigr).
 \label{JRod}\end{align}

{
}
\end{document}